\documentclass{article}[11pt]  
\usepackage{amsmath}
\usepackage{amssymb}
\usepackage{amsthm}
\usepackage{ifpdf}
\usepackage{wrapfig}
\usepackage{fullpage}
\usepackage{cite}
\usepackage{subcaption}
\usepackage{float}
\usepackage{enumitem}
\usepackage{graphicx}
\usepackage[noend]{algorithm2e}
\usepackage{multirow}
\usepackage[table]{xcolor}
\usepackage[hidelinks]{hyperref}

\usepackage{authblk}

\theoremstyle{definition}
\newtheorem{theorem}{Theorem}[section]
\newtheorem{lemma}[theorem]{Lemma}
\newtheorem{corollary}[theorem]{Corollary}

\newtheorem{definition}[theorem]{Definition}

\definecolor{highlight}{HTML}{FFDD6B}
\definecolor{greyh}{HTML}{DDDDDD}

\newcommand{\BO}[1]{\mathcal{O}(#1)}

\title{Building Squares with Optimal State Complexity in Restricted Active Self-Assembly\footnote{This research was supported in part by National Science Foundation Grant CCF-1817602.
}}

\author[1]{Robert M. Alaniz}
\author[1]{David Caballero}
\author[1]{Sonya C. Cirlos}
\author[2]{Timothy Gomez}
\author[1]{Elise Grizzell}
\author[1]{Andrew Rodriguez}
\author[1]{Robert Schweller}
\author[1]{Armando Tenorio}
\author[1]{Tim Wylie}

\affil[1]{University of Texas Rio Grande Valley, Edinburg, TX, USA.\\ \{robert.alaniz01,david.caballero01,sonya.cirlos01,elise.grizzell01\}@utrgv.edu,
\{andrew.rodriguez09,robert.schweller,armando.tenorio01,timothy.wylie\}@utrgv.edu}
\affil[2]{Massachusetts Institute of Technology, Cambridge, MA, USA.\\ tagomez7@mit.edu}

\begin{document}

\date{}
\clearpage\maketitle
\thispagestyle{empty}

\vspace*{-.5cm}
\begin{abstract}
Tile Automata is a recently defined model of self-assembly that borrows many concepts from cellular automata to create active self-assembling systems where changes may be occurring within an assembly without requiring attachment. This model has been shown to be powerful, but many fundamental questions have yet to be explored. Here, we study the state complexity of assembling $n \times n$ squares in seeded Tile Automata systems where growth starts from a seed and tiles may attach one at a time, similar to the abstract Tile Assembly Model. We provide optimal bounds for three classes of seeded Tile Automata systems (all without detachment), which vary in the amount of complexity allowed in the transition rules. We show that, in general, seeded Tile Automata systems require $\Theta{(\log^{\frac{1}{4}} n)}$ states. For single-transition systems, where only one state may change in a transition rule, we show a bound of $\Theta{(\log^{\frac{1}{3}} n)}$, and for deterministic systems, where each pair of states may only have one associated transition rule, a bound of $\Theta( (\frac{\log n}{\log \log n})^\frac{1}{2} )$. 
\end{abstract}

\newpage
\tableofcontents
\newpage
\section{Introduction}

Self-assembly is the process by which simple elements in a system organize themselves into more complex structures based on a set of rules that govern their interactions. These types of systems occur naturally and can be easily constructed artificially to offer many advantages when building micro or nanoscale objects. With many ways to create self-assembling systems, new models of abstraction have also arisen to handle specific mechanisms and procedures. These models continue to evolve and be extended as technology advances, and it becomes important to connect and relate different models through this process to understand the capabilities of each type of system.

In the abstract Tile Assembly Model (aTAM)  \cite{Winf98}, the elements of a system are represented using labeled unit squares called tiles. A system is initialized with a seed (a tile or assembly) that grows as other single tiles attach until there are no more valid attachments. The behavior of a system can then be programmed using the interactions of tiles, and is known to be capable of Turing Computation \cite{Winf98}, is Intrinsically Universal \cite{doty2012tile}, and can assemble general scaled shapes \cite{soloveichik2007complexity}. However, many of these results utilize a concept called \emph{cooperative binding}, where a tile must attach to an assembly using the interaction from two other tiles. Unlike with cooperative binding, the non-cooperative aTAM is not Intrinsically Universal \cite{reqCoop, t1notIU} and more recent work has shown that it is not capable of Turing Computation \cite{decidableATAM}. Many extensions of this model increase the power of non-cooperative systems \cite{duples, negt1, stepwise, fu2012self, hader2021geometric, gilbert2016computing}.

One recent model of self-assembly is Tile Automata \cite{freezeSim}. This model marries the concept of state changes from Cellular Automata \cite{ca110, worsch2013towards, goles2011communication} and the assembly process from the 2-Handed Assembly model (2HAM) \cite{2HABTO}. 
Previous work \cite{freezeSim, taAmb, cantu2020signal} explored Tile Automata as a unifying model for comparing the relative powers of different Tile Assembly models. The complexity of verifying the behavior of systems, along with their computational power, was studied in \cite{veriTA}.
Many of these works impose additional experimentally motivated limitations on the Tile Automata model that help connect the model and its capabilities to potential molecular implementations, such as using DNA assemblies with sensors to assemble larger structures \cite{green2019autonomous}, building spatial localized circuits on DNA origami \cite{chatterjee2017a}, or DNA walkers that sort cargo \cite{cargo}.

In this paper, we explore the aTAM generalized with state changes; we define our producible assemblies as what can be grown by attaching tiles one at a time to a seed tile or performing transition rules, which we refer to as seeded Tile Automata.
This is a bounded version of Asynchronous Cellular Automata \cite{fates2013guided}. Reachability problems, which are similar to verification problems in self-assembly, have been studied with many completeness results \cite{dennunzio2017computational}. Further, the freezing property used in this and previous work also exists in Cellular Automata \cite{goles2015introducing, ollinger2019freezing}.\footnote{We would like to thank a reviewer for bringing these works to our attention. }
Freezing is defined differently in Cellular Automata by requiring that there exists an ordering to the states.

While Tile Automata has many possible metrics, we focus on the number of states needed to uniquely assemble $n \times n$ squares at the smallest constant temperature, $\tau = 1$.
We achieve optimal bounds in three versions of the model with varying restrictions on the transition rules. Our results, along with previous results in the aTAM, are outlined in Table \ref{tab:table}.

\begin{table}[t]
    \centering 
    \renewcommand{\arraystretch}{1.2}
    \begin{tabular}[b]{ | c | c | c | c | c | c | c }
        \hline
         \multirow{2}{*}{\textbf{Model}} & \multirow{2}{*}{\textbf{$\tau$}} & \multicolumn{3}{c|}{ $n \times n$  \textbf{Squares} }  \\  \cline{3-5}
         
         & & \textbf{Lower} &  \textbf{Upper} & \textbf{Theorem}  \\ \hline 
        aTAM & $1$  & $\Omega(\frac{\log n}{\log\log n})$ & $\BO{n}$ &   \cite{rothemund2000program}, \cite{adleman2001running} \\ \hline
        
        aTAM & $2$ & \multicolumn{2}{c|}{$\Theta(\frac{\log n}{\log\log n})$} &  \cite{rothemund2000program},\cite{adleman2001running}  \\ \hline
        
        Flexible Glue aTAM & $2$ & \multicolumn{2}{c|}{$\Theta(\log ^\frac{1}{2} n)$} & \cite{gen2005}  \\ \hline
        
        Seeded TA Det. & $1$ &  \multicolumn{2}{c|}{$\Theta( (\frac{\log n}{\log \log n})^\frac{1}{2} )$}  &  Thms. \ref{thm:determ}, \ref{thm:SQ} \\ \hline
        
        Seeded TA ST & $1$ & \multicolumn{2}{c|}{ $\Theta(\log ^\frac{1}{3} n)$} & Thms. \ref{thm:srLB}, \ref{thm:SQ}   \\ \hline
        
        Seeded TA & $1$ &   \multicolumn{2}{c|}{$\Theta(\log ^\frac{1}{4} n)$}  & Thms. \ref{thm:ndLB}, \ref{thm:SQ}   \\ \hline
    \end{tabular}
    \caption{Bounds on the number of states for $n \times n$ squares in the Abstract Tile Assembly model, with and without cooperative binding, and the seeded Tile Automata model with our transition rules. ST stands for single-transition.}
    \label{tab:table}
\end{table}

\begin{table}[t]
    \centering 
    \renewcommand{\arraystretch}{1.2}
    \begin{tabular}[b]{ | c | c | c | c | c | c | c }
        \hline
         \multirow{2}{*}{\textbf{Model}}  & \multicolumn{3}{c|}{ \textbf{Length-}$n$  \textbf{Binary Strings} }  \\  \cline{2-4}
         
         & \textbf{Lower} &  \textbf{Upper} & \textbf{Theorem}  \\ \hline 
        
        Seeded TA ST Det. &   \multicolumn{2}{c|}{$\BO{n^\frac{1}{2} }$}  &  Thms. \ref{thm:detStringLB}, \ref{thm:2LU} \\ \hline
        
        Seeded TA Det. &  \multicolumn{2}{c|}{$\Theta( \frac{n}{\log n}^\frac{1}{2} )$}  &  Thm. \ref{thm:detStringLB}, \ref{thm:optDetStr}  \\ \hline
        
        Seeded TA ST &  \multicolumn{2}{c|}{ $\Theta(n ^\frac{1}{3})$} & Thms. \ref{thm:STStringLB}, \ref{thm:SRstrings}   \\ \hline
        
        Seeded TA &  \multicolumn{2}{c|}{$\Theta(n ^\frac{1}{4})$}  & Thms. \ref{thm:ndStringLB}, \ref{thm:nonDetstrings}   \\ \hline
    \end{tabular}
    \caption{Bounds on the number of states for construction length-$n$ strings.}
    \label{tab:stringTable}
\end{table}

%
%
%
%
%

\begin{table}[t]
    \centering 
    \renewcommand{\arraystretch}{1.2}
    \begin{tabular}[b]{ | c | c | @{}c@{} | c | c | c | c | c |}
        \hline
         \multirow{2}{*}{\textbf{Model}} & \multirow{2}{*}{\textbf{$\tau$}}  & \multirow{2}{*}{\textbf{Height}} & \multicolumn{3}{c|}{ $k \times n$  \textbf{Rectangles} }  \\  \cline{4-6}
         
         & & & \textbf{Lower} &  \textbf{Upper} & \textbf{Theorem}  \\ \hline 
        aTAM & $1$ &  $k$ & $\Omega(n^\frac{1}{k})$ & $\BO{n}$ & \cite{furcy2019new},  \cite{adleman2001running}\\ \hline
        
        aTAM & $2$ & $k$ & $\Omega(\frac{n^{\frac{1}{k}}}{k})$  & $\BO{n^{\frac{1}{k}}}$ & \cite{rothemund2000program},  \cite{gen2005}  \\ \hline
        
        Flexible Glue aTAM & $2$ & $k$ &  $\Omega(\frac{n^{\frac{1}{k}}}{k})$  & $\BO{n^{\frac{1}{k}}}$  &  \cite{gen2005}  \\ \hline
        
        Seeded TA Det. & $1$ & $1$ &  \multicolumn{2}{c|}{$\Theta( (\frac{\log n}{\log \log n})^\frac{1}{2} )$}  &  Thms. \ref{thm:determ}, \ref{thm:detConstHeight} \\ \hline
        
        Seeded TA Det. ST & $1$ & $2$ &  \multicolumn{2}{c|}{$\Theta( (\frac{\log n}{\log \log n})^\frac{1}{2} )$}  &  Thms. \ref{thm:determ}, \ref{thm:detConstHeightScaled} \\ \hline
        
        Seeded TA ST & $1$ & $3$ & \multicolumn{2}{c|}{ $\Theta(\log ^\frac{1}{3} n)$} & Thms. \ref{thm:srLB}, \ref{thm:STconstHeight}  \\ \hline
        
        Seeded TA & $1$ & $4$  &  \multicolumn{2}{c|}{$\Theta(\log ^\frac{1}{4} n)$}  & Thms. \ref{thm:ndLB}, \ref{thm:constHeight}   \\ \hline
    \end{tabular}
    \caption{Lower and upper complexity bounds on the number of states for constructing $k \times n$ thin rectangles for $k \leq \frac{\log n}{\log \log n - \log \log \log n}$. $\tau$ is the temperature of the system.}
    \label{tab:rectTable}
\end{table}

\subsection{Previous Work}
In the aTAM, the number of tile types needed, for nearly all $n$, to construct an $n \times n$ square is $\Theta(\frac{\log n }{ \log n \log n})$ \cite{rothemund2000program, adleman2001running} with temperature $\tau = 2$ (row 2 of Table \ref{tab:table}). The same lower bounds hold for $\tau = 1$ (row 1 of Table \ref{tab:table}). The run time of this system was also shown to be optimal $\Theta(n)$ \cite{adleman2001running}. Other bounds for building rectangles were shown in \cite{gen2005}. While no tighter bounds\footnote{Other than trivial $\BO{n}$ bounds.} have been shown for $n \times n$ squares at $\tau = 1$ in the aTAM, generalizations to the model that allow (just-barely) 3D growth have shown an upper bound of $\BO{\log n}$ for tile types needed \cite{temp13D}. Recent work in \cite{dna27ThinRect} shows improved upper and lower bounds on building thin rectangles in the case of $\tau = 1$ and in (just-barely) 3D.

Other models of self-assembly have also been shown to have a smaller tile complexity, such as the staged assembly model \cite{demaine2008staged, chalk2018optimal} and temperature programming \cite{tempProgram}.
Investigation into different active self-assembly models have also explored the run time of systems \cite{woods2013active, runTimeCRNTAM}.

\subsection{Our Contributions}
In this work, we explore building an important benchmark shape, squares, in $\tau = 1$ seeded Tile Automata. The techniques presented also provide results for other benchmark problems in strings and constant-height rectangles. We consider only affinity-strengthening transition rules that remove the ability for an assembly to break apart. Our results are shown in Tables \ref{tab:stringTable}, \ref{tab:table} and \ref{tab:rectTable}.

We start in Section \ref{sec:lowerBounds} by proving lower bounds for building $m \times n$ rectangles ($m < n$) 
based on three different transition rule restrictions. The first is nondeterministic or general seeded Tile Automata, where there are no restrictions and a pair of states may have multiple transition rules. The second is single-transition rules where only one tile may change states in a transition rule, but we still allow multiple rules for each pair of states. The last restriction, deterministic, is the most restrictive where each pair of states may only have one transition rule (for each direction). 

In Section \ref{sec:strings}, we use transition rules to optimally encode strings in the various versions of the model. We start with two deterministic constructions that both achieve the optimal bound. The difference between the two constructions is the first is 2D with single-transitions and the second is 1D with double-transitions. We extend these constructions in Section \ref{sec:nonDetS} to use nondeterministic transition rules to achieve more efficient bounds. 
  
Section \ref{sec:rectangles} contains a method to build a base-$b$ counter similar to previous work but using only $\tau = 1$ transition rules. This can grow off the previous construction by encoding a binary string to build length-$n$ rectangles where each tile only transitions a constant number of times. Additionally, we provide an alternate construction that can build constant-height length-$n$ rectangles. 

Finally, we present our construction for building squares in Section \ref{sec:squares} by applying the methods from rectangle building. We achieve optimal state complexity for each kind of transition rules mentioned.

This paper is an extension of a conference publication \cite{Alaniz:2022:SAND}, with additional content for the benchmark problems of building strings and rectangles. The lower bound proofs have been generalized to arbitrary rectangles and strings, and we give constructions with matching complexity. The new upper bounds to build $1$-dimensional strings are given in subsection \ref{sec:strings}. Further, we give constructions to build optimal constant height rectangles. 


\textbf{AutoTile.}
To test our constructions, we developed AutoTile, a seeded Tile Automata simulator. Each system discussed in the paper is currently available for simulation. 
AutoTile is available at \url{https://github.com/asarg/AutoTile}.


\section{Definitions}
\begin{figure}[t]
	\centering
	\begin{subfigure}[b]{.27\textwidth}
		\centering
		\includegraphics[width=.9\textwidth]{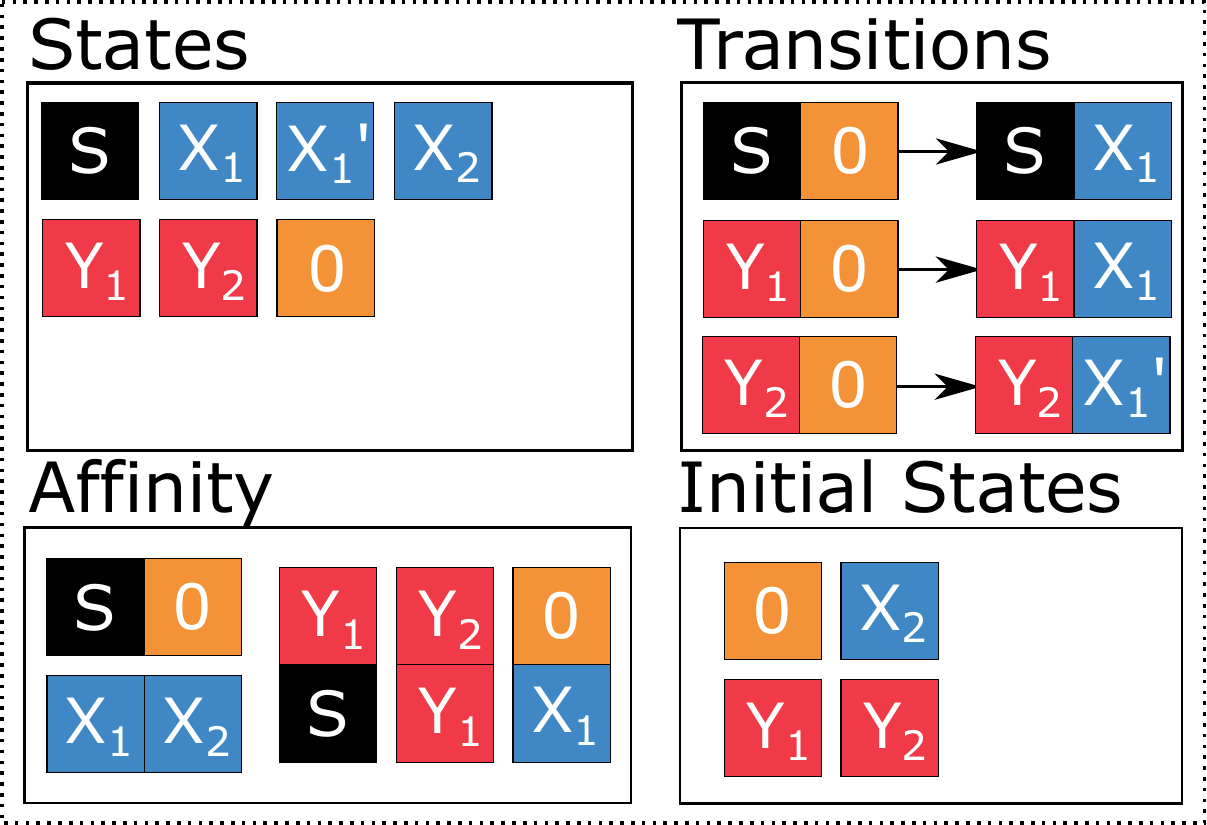}
		\caption{}
		\label{fig:SquareExampleSystem}
	\end{subfigure}
	\begin{subfigure}[b]{.7\textwidth}
		\centering
		\includegraphics[width=1.\textwidth]{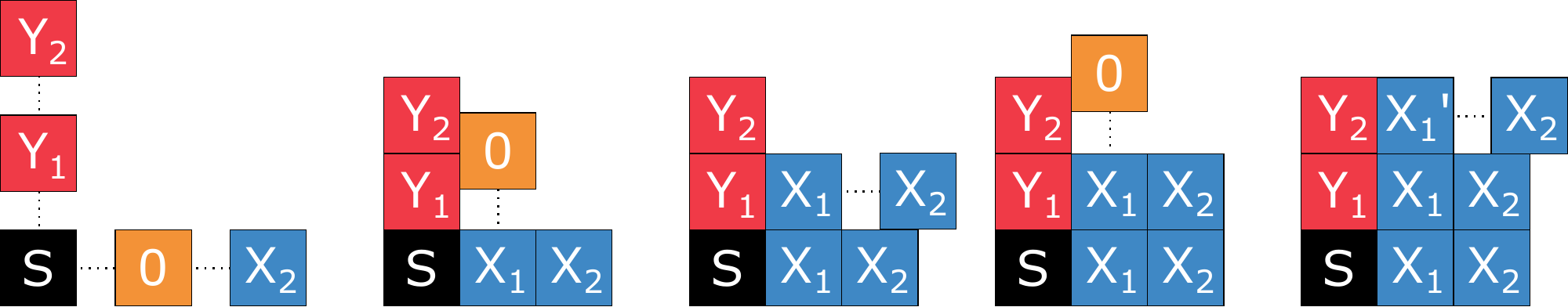}
		\caption{}
		\label{fig:SquareExample}
	\end{subfigure}
    \caption{(a) Example of a Tile Automata system, it should be noted that $\tau = 1$ and state $S$ is our seed. (b) A walk-through of our example Tile Automata system building the $3 \times 3$ square it uniquely produces. We use dotted lines throughout our paper to represent tiles attaching to one another.  } \label{fig:exta}
\end{figure}

The Tile Automata model differs quite a bit from normal self-assembly models since a \emph{tile} may change \emph{state}, which draws inspiration from Cellular Automata. Thus, there are two aspects of a TA system being: the self-assembling that may occur with tiles in a state and the changes to the states once they have attached to each other. To address these aspects, we define the building blocks and interactions, and then the definitions around the model and what it may assemble or output. Finally, since we are looking at a limited TA system, we also define specific limitations and variations of the model. For reference, an example system is shown in Figure \ref{fig:exta}.

\subsection{Building Blocks}
The basic definitions of all self-assembly models include the concepts of tiles, some method of attachment, and the concept of aggregation into larger assemblies. The Cellular Automata aspect also brings in the concept of transitions.

\textbf{Tiles.} Let $\Sigma$ be a set of \emph{states} or symbols. A tile $t = (\sigma, p)$ is a non-rotatable unit square placed at point $p \in \mathbb{Z}^2$ and has a state of $\sigma \in \Sigma$.


\textbf{Affinity Function.} An \emph{affinity function} $\Pi$ over a set of states $\Sigma$ takes an ordered pair of states $(\sigma_1, \sigma_2) \in \Sigma \times \Sigma$ and an orientation $d \in D$, where $D = \{\perp,\vdash\}$, and outputs an element of $\mathbb{Z}^0$. The orientation $d$ is the relative position to each other with $\vdash$  meaning horizontal and $\perp$ meaning vertical, with the $\sigma_1$  being the west or north state respectively.  We refer to the output as the \emph{Affinity Strength} between these two states.

\textbf{Transition Rules.}
A \emph{Transition Rule} consists of two ordered pairs of states $(\sigma_1, \sigma_2), (\sigma_3, \sigma_4)$ and an orientation $d \in D$, where $D = \{\perp,\vdash\}$. This denotes that if the states  $(\sigma_1, \sigma_2)$ are next to each other in orientation $d$ ($\sigma_1$ as the west/north state) they may be replaced by the states $(\sigma_3, \sigma_4)$.

\textbf{Assembly.} An assembly $A$ is a set of tiles with states in $\Sigma$ such that for every pair of tiles $t_1 = (\sigma_1, p_1), t_2 = ( \sigma_2, p_2)$, $p_1 \neq p_2$. Informally, each position contains at most one tile.
Further, we say assemblies are equal in regards to translation. Two assemblies $A_1$ and $A_2$ are equal if there exists a vector $\vec{v}$ such that $A_1 = A_2 + \vec{v}$.

Let $B_G(A)$ be the bond graph formed by taking a node for each tile in $A$ and adding an edge between neighboring tiles $t_1 = (\sigma_1, p_1)$ and $t_2 = (\sigma_2, p_2)$ with a weight equal to $\Pi(\sigma_1, \sigma_2)$.
We say an assembly $A$ is $\tau$-stable for some $\tau \in \mathbb{Z}^0$ if the minimum cut through $B_G(A)$ is greater than or equal to $\tau$.

\subsection{The Tile Automata Model}
Here, we define and investigate the \emph{Seeded Tile Automata} model, which differs by only allowing single tile attachments to a growing seed similar to the aTAM.

\textbf{Seeded Tile Automata.}
A Seeded Tile Automata system is a 6-tuple $\Gamma = ( \Sigma, \Lambda, \Pi, \Delta, s, \tau )$ where $\Sigma$ is a set of states, $\Lambda \subseteq \Sigma$ a set of \emph{initial states}, $\Pi$ is an \emph{affinity function}, $\Delta$ is a set of \emph{transition rules},  $s$ is a stable assembly called the \emph{seed} assembly, and $\tau$ is the \emph{temperature} (or threshold).  Our results use the most restrictive version of this model where $s$ is a single tile.

\textbf{Attachment Step.}
A tile $t = (\sigma, p)$ may attach to an assembly $A$ at temperature $\tau$ to build an assembly $A' = A \bigcup t$ if $A'$ is $\tau$-stable and $\sigma \in \Lambda$. We denote this as $A \rightarrow_{\Lambda, \tau} A'$.

\textbf{Transition Step.}
An assembly $A$ is transitionable to an assembly $A'$ if there exists two neighboring tiles $t_1 = (\sigma_1, p_1), t_2 = (\sigma_2, p_2) \in A$ (where $t_1$ is the west or north tile) such that there exists a transition rule in $\Delta$ with the first pair being $(\sigma_1, \sigma_2)$ and $A' = (A \setminus \{t_1, t_2\} ) \bigcup \{ t_3 = (\sigma_3, p_1), t_4 = (\sigma_4, p_2)\}$. We denote this as $A \rightarrow_\Delta A'$.

\textbf{Producibles.}
We refer to both attachment steps and transition steps as production steps, we define  $A \rightarrow_* A'$ as the transitive closure of $A \rightarrow_{\Lambda, \tau} A'$ and $A \rightarrow_\Delta A'$.
The set of \emph{producible assemblies} for a Tile Automata system $\Gamma = ( \Sigma, \Lambda, \Pi, \Delta, s, \tau )$ is written as $PROD(\Gamma)$. We define $PROD(\Gamma)$ recursively as follows,
\begin{itemize}
	\item $s \in PROD(\Gamma)$
	\item $A' \in PROD(\Gamma)$ if $\exists A \in PROD(\Gamma)$ such that $A \rightarrow_{\Lambda, \tau} A'$.
	\item $A' \in PROD(\Gamma)$ if $\exists A \in PROD(\Gamma)$ such that $A \rightarrow_{\Delta} A'$.
\end{itemize}

\textbf{Terminal Assemblies.}
The set of terminal assemblies for a Tile Automata system $\Gamma = (\Sigma, \Lambda, \Pi, \Delta, \tau)$ is written as $TERM(\Gamma)$. This is the set of assemblies that cannot grow or transition any further. Formally, an assembly $A \in TERM(\Gamma)$ if $A \in PROD(\Gamma)$ and there does not exists any assembly $A' \in PROD(\Gamma)$ such that  $A \rightarrow_{\Lambda, \tau} A'$ or $A \rightarrow_{\Delta} A'$.
A Tile Automata system $\Gamma = ( \Sigma, \Lambda, \Pi, \Delta, s, \tau )$ \emph{uniquely} assembles an assembly $A$ if $A \in TERM(\Gamma)$, and for all $A' \in PROD(\Gamma), A' \rightarrow_* A$.

\subsection{Limited Model Reference}

We explore an extremely limited version of seeded TA that is affinity-strengthening, freezing, and may be a single-transition system. We investigate both deterministic and nondeterministic versions of this model.


\textbf{Affinity Strengthening.}
We only consider transitions rules that are affinity strengthening, meaning for each transition rule $((\sigma_1, \sigma_2), (\sigma_3, \sigma_4), d)$, the bond between $(\sigma_3, \sigma_4)$ must be at least the strength of $(\sigma_1, \sigma_2)$. Formally, $\Pi(\sigma_3, \sigma_4, d) \geq \Pi(\sigma_1, \sigma_2, d)$. This ensures that transitions may not induce cuts in the bond graph.

In the case of non-cooperative systems ($\tau = 1$), the affinity strength between states is always $1$ so we may refer to the affinity function as an affinity set $\Lambda_s$, where each affinity is a $3$-pule $(\sigma_1, \sigma_2, d)$.

\textbf{Freezing.}
Freezing systems were introduced with Tile Automata. A freezing system simply means that a tile may transition to any state only once. Thus, if a tile is in state $A$ and transitions to another state, it is not allowed to ever transition back to $A$.

\textbf{Deterministic vs. Nondeterministic.}
For clarification, a deterministic system in TA has only one possible production step at a time, whether that be an attachment or a state transition. A nondeterministic system may have many possible production steps and any choice may be taken.

\textbf{Single-Transition System.}
We restrict our TA system to only use single-transition rules. This means that for each transition rule one of the states may change, but not both. Note that nondeterminism is still allowed in this system.

\begin{figure}[t]
	\centering
	\begin{subfigure}[b]{.23\textwidth}
		\centering
		\includegraphics[width=.9\textwidth]{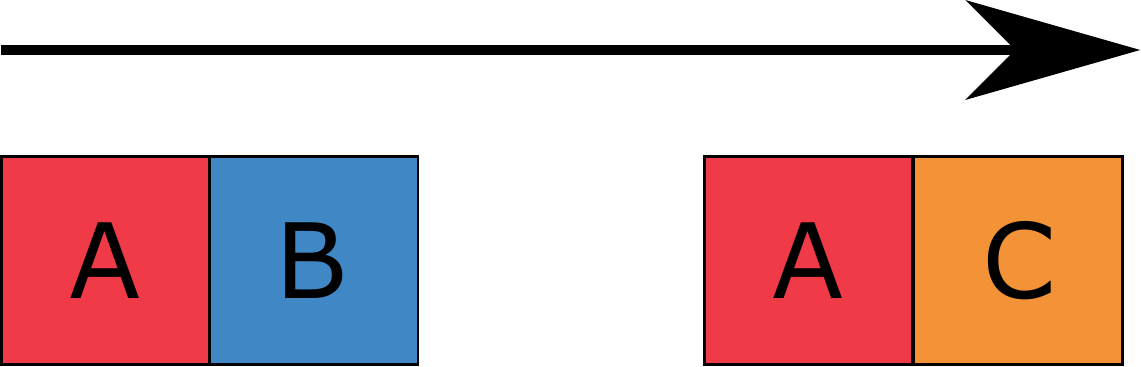}
		\caption{}
		\label{fig:detSR}
	\end{subfigure}
	\begin{subfigure}[b]{.23\textwidth}
		\centering
		\includegraphics[width=.9\textwidth]{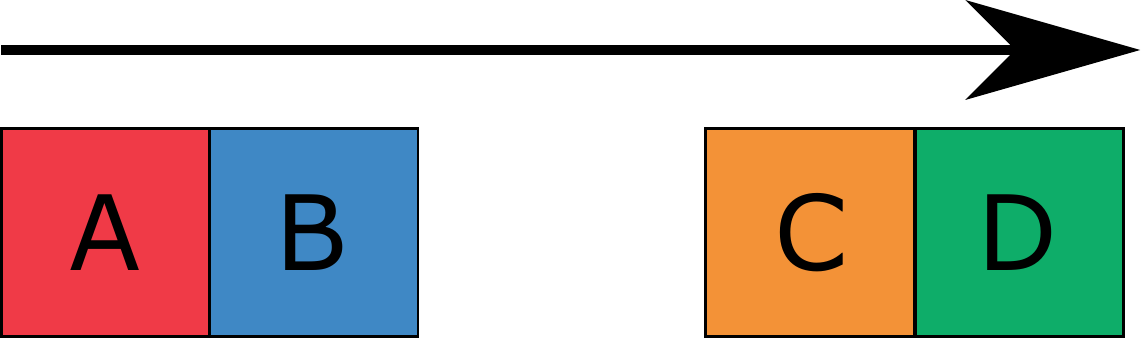}
		\caption{}
		\label{fig:detDR}
	\end{subfigure}
	\begin{subfigure}[b]{.23\textwidth}
		\centering
		\includegraphics[width=.9\textwidth]{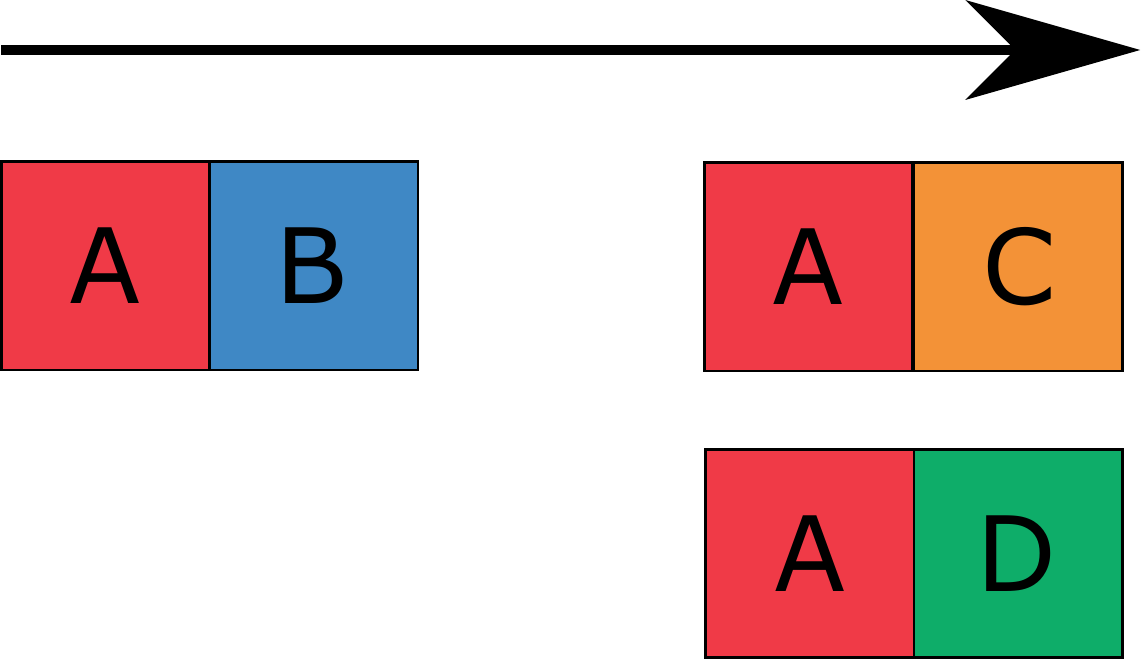}
		\caption{}
		\label{fig:ndSR}
	\end{subfigure}
	\begin{subfigure}[b]{.23\textwidth}
		\centering
		\includegraphics[width=.9\textwidth]{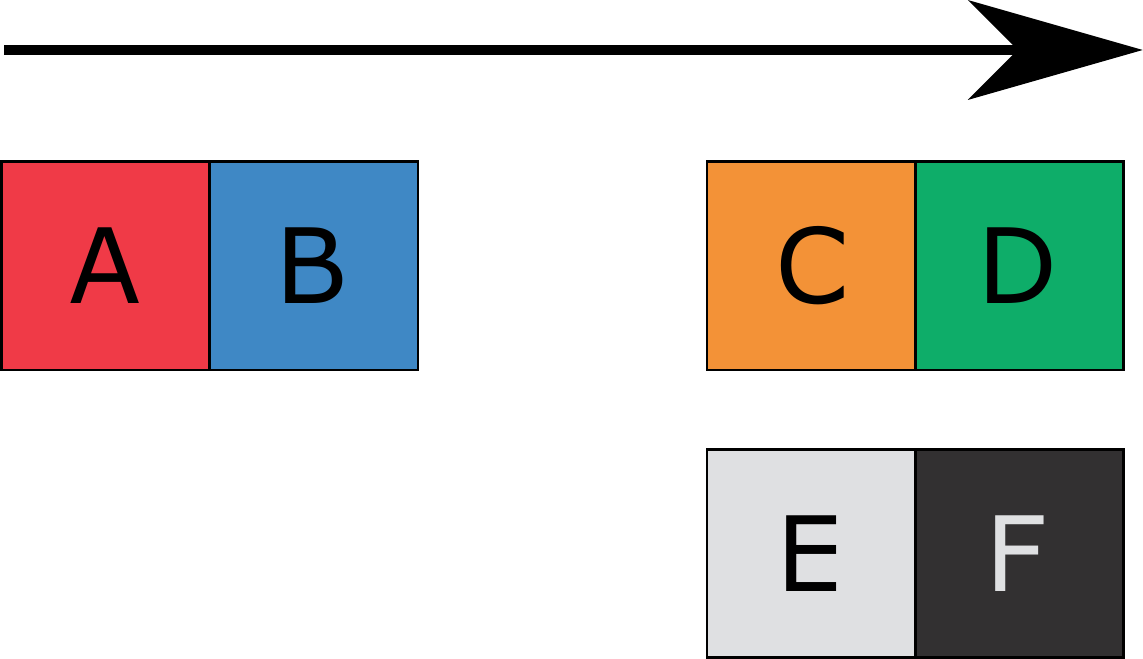}
		\caption{}
		\label{fig:ndDR}
	\end{subfigure}
    \caption{Example of a (a) Deterministic single-transition rule (b) Deterministic transition rule (c) Single-Transition rule (d) Nondeterministic transition rule. Note that the deterministic rules can only transition that pair of tiles to one other possible state, while nondeterministic rules can transition to multiple possible states. }
\end{figure}

%

\section{State Space Lower Bounds}\label{sec:lowerBounds}


Let $p(n)$ be a function from the positive integers to the set $\{0,1\}$, informally termed a \emph{proposition}, where $0$ denotes the proposition being false and $1$ denotes the proposition being true.  We say a proposition $p(n)$ holds for \emph{almost all} $n$ if $\lim_{n \to \infty} \frac{1}{n}\sum_{i=1}^n p(i) = 1$.


\begin{lemma}\label{lemma:lowerBits}
Let $U$ be a set of TA systems, $b(\cdot)$ be an injective function mapping each element of $U$ to a string of bits, and $\epsilon$ a real number from $0 < \epsilon < 1$.  Then for almost all integers $n$, any TA system $\Gamma \in U$ that uniquely assembles an $n \times m$ rectangle for $n \geq m$ has a bit-string of length $|b(\Gamma)| \geq (1-\epsilon)\log n$.
\end{lemma}

\begin{proof}
For a given $i \geq j \geq 1$, let $M_i \in U$ denote the TA system in $U$ with the minimum value $|b(M_i)|$ over all systems in $U$ that uniquely assemble an $i\times j$ rectangle, and let $M_i$ be undefined if no such system in $U$ builds such a shape. Let $p(i)$ be the proposition that $|b(M_i)| \geq (1-\epsilon)\log i$.  We show that $\lim_{n\to\infty} \frac{1}{n}\sum_{i=1}^{n} p(i) = 1$.  Let $R_n = \{ M_i | 1\leq i\leq n, |b(M_i)| < (1-\epsilon)\log n \}$.  Note that $n-|R_n| \leq \sum_{i=1}^{n} p(i)$. By the pigeon-hole principle, $|R_n|\leq 2^{(1-\epsilon)\log n} = n^{(1-\epsilon)}$. Therefore,

\[\lim_{n\to\infty} \frac{1}{n}\sum_{i=1}^{n} p(i) \geq \lim_{n\to\infty} \frac{1}{n}(n-|R_n|) \geq \lim_{n\to\infty} \frac{1}{n}(n- n^{1-\epsilon}) = 1.\]
\end{proof}


\begin{theorem}[Deterministic TA]\label{thm:determ}
For almost all $n$, any deterministic Tile Automata system that uniquely assembles an $n \times m$ rectangle with $n \geq m$ contains $\Omega{(\frac{\log n}{\log\log n})^\frac{1}{2}}$ states.
\end{theorem}

\begin{proof}
We create an injective mapping $b(\Gamma)$ from any deterministic TA system to bit-strings in the following manner.  Let $\Sigma$ denote the set of states in a given system.  We encode the state set in $\BO{\log |\Sigma|}$ bits, we encode the affinity function in a $|\Sigma|\times |\Sigma|$ table of strengths in $\BO{|\Sigma|^2}$ bits (assuming a constant bound on bonding thresholds), and we encode the rules of the system in an $|\Sigma|\times |\Sigma|$ table mapping pairs of rules to their unique new pair of rules using $\BO{|\Sigma|^2 \log |\Sigma|}$ bits, for a total of $\BO{|\Sigma|^2\log |\Sigma|}$ bits to encode any $|\Sigma|$ state system.

Let $\Gamma_n$ denote the smallest state system that uniquely assembles an $n \times m$ rectangle, and let $\Sigma_n$ denote the state set.  By Lemma~\ref{lemma:lowerBits}, $|b(\Gamma_n)| \geq (1-\epsilon)\log n$ for almost all $n$, and so $|\Sigma_n|^2\log |\Sigma_n| = \Omega(\log n)$ for almost all $n$.  We know that $|\Sigma_n| = \BO{\log n}$, so for some constant $c$, $|\Sigma_n| \geq c (\frac{\log n}{\log\log n})^\frac{1}{2}$ for almost all $n$.
\end{proof}

\begin{theorem}[Nondeterministic TA]\label{thm:ndLB}
For almost all $n$, any Tile Automata system (nondeterministic) that uniquely assembles an $n \times m$ rectangle with $n \geq m$ contains $\Omega{(\log^{\frac{1}{4}} n)}$ states.
\end{theorem}
\begin{proof}
We create an injective mapping $b(\Gamma)$ from nondeterministic TA systems to bit-strings.  We use the same mapping as in Theorem~\ref{thm:determ} except for the rule encoding, and now use a $|\Sigma|^4$ binary table to specify which rules are, or are not, present in the system.  From Lemma~\ref{lemma:lowerBits},  the minimum state system to build an $n \times m$ rectangle has $|\Sigma_n|^4 = \Omega(\log n)$ states for almost all $n$, which implies that $|\Sigma_n| = \Omega(\log^{\frac{1}{4}} n)$ for almost all $n$.
\end{proof}

\begin{theorem}[Single-Transition TA]\label{thm:srLB}
For almost all $n$, any single-transition Tile Automata system (nondeterministic) that uniquely assembles an $n \times m$ rectangle with $n \geq m$ contains $\Omega{(\log^{\frac{1}{3}} n)}$ states.
\end{theorem}
\begin{proof}
We use the same bit-string encoding $b(\Gamma)$ as in Theorem~\ref{thm:determ}, except for the encoding of the ruleset, and we use a $|\Sigma|^2\times |\Sigma|$ matrix of constants.  The encoding works by encoding the pair of states for a transition in the first column, paired with the second column dictating the state that will change, and the entry in the table denoting which of the two possible states changed.  We thus encode an $|\Sigma|$-state system in $\BO{|\Sigma|^3}$ bits, which yields $|\Sigma| = \Omega(\log^\frac{1}{3} n)$ for almost all $n$ from Lemma~\ref{lemma:lowerBits}.
\end{proof}

\subsection{Strings}
Here, we provide lower bounds on building binary strings. These bounds hold at any scale. We define scaled strings using macroblocks. A macroblock is an $a \times b$ subassembly that maps to a specific bit of the string. Let $M^{a,b}(\Sigma)$ be the set of all size $a \times b$ macroblocks over states $\Sigma$. Let the $i^{th}$ macroblock of an assembly be the size $a \times b$ subassembly whose lower left tile is at location $(a\cdot i, 0)$ in the plane.

\begin{definition}[String Representation]
An assembly $A$, over states $\Sigma$, represents a string $S$ over a set of symbols $L$ at scale $a \times b$, if there exists a mapping from the elements of $M^{a,b}(\Sigma)$ to the elements of $L$, and the $i^{th}$ macroblock of $A$ maps to the $i^{th}$ symbol of $S$.
\end{definition}


\begin{lemma}\label{lem:stringMap}
Let $U$ be a set of TA systems and $b$ be an injective function mapping each element of $U$ to a string of bits. Then for all $n \geq 0$, there exists a string $S$ of length $n$ such that any TA system $\Gamma \in U$ that uniquely assembles an assembly $A$ that represents $S$ at any scale has $|b(\Gamma)| \geq n$.
\end{lemma}
\begin{proof}
Given the string $b(\Gamma)$, the assembly $A$ can be computed and the string $S$ output after reading from each macroblock.
There are $2^n$ length-$n$ strings, but only $2^n - 1$ bit-strings with size less than $n$, so by the pigeonhole principle, at least one of the systems must map to a string of length $n$.
\end{proof}

\begin{theorem}[Deterministic TA]\label{thm:detStringLB}
For all $n > 0$, there exists a binary string $S$ of length $n$, such that any deterministic Tile Automata that uniquely assembles an assembly that represents $S$ at any scale contains $\Omega( \frac{n}{\log |n|}^\frac{1}{2} )$ states.
\end{theorem}
\begin{proof}
Using the encoding method described in Theorem \ref{thm:determ}, we represent a TA system with $|\Sigma|$ states in $\BO{|\Sigma|^2 \log |\Sigma|}$ bits.
By Lemma \ref{lem:stringMap}, we know there exists a length-$n$ binary string $S$ such that any system $\Gamma$ that uniquely assembles $S$ requires $n$ bits to describe.

\begin{align}
\BO{|\Sigma|^2 \log |\Sigma|} &\geq n\\
 \BO{|\Sigma|^2} &\geq \frac{n}{\log |\Sigma|}\\
 |\Sigma| &\geq \Omega ( \frac{n}{\log |\Sigma|}^\frac{1}{2} )
\end{align}
A trivial construction of assigning each bit a unique state, gives $|\Sigma| \leq n$.
\[ |\Sigma| \geq \Omega ( \frac{n}{\log n}^\frac{1}{2} ) \]
\end{proof}

\begin{theorem}[Single-Transition TA]\label{thm:STStringLB}
For all $n > 0$, there exists a binary string $S$ of length $n$, such that any single-transition Tile Automata that uniquely assembles an assembly that represents $S$ at any scale contains $\Omega(n^\frac{1}{3})$ states.
\end{theorem}
\begin{proof}
With Theorem \ref{thm:srLB}, we may encode the system with $\BO{\Sigma^3}$ bits and Lemma \ref{lem:stringMap} implies there exists a string where $\BO{\Sigma^3} \geq n$, and thus $|\Sigma| \geq \Omega(n^\frac{1}{3})$.
\end{proof}

\begin{theorem}[Nondeterministic TA]\label{thm:ndStringLB}
For all $n > 0$, there exists a binary string $S$ of length $n$, such that any Tile Automata (in particular any nondeterministic system) that uniquely assembles an assembly that represents $S$ at any scale contains $\Omega(n^\frac{1}{4})$ states.
\end{theorem}

\begin{proof}
The encoding method from Theorem \ref{thm:ndLB} and bound from Lemma \ref{lem:stringMap} give us a lower bound of $\Omega(n^\frac{1}{4})$ states.
\end{proof}

\section{String Unpacking}\label{sec:strings}
A key tool in our constructions is the ability to build strings efficiently. We do so by encoding the string in the transition rules. We first describe the simplest version of this construction that has freezing deterministic single-transition rules achieving $\mathcal{O}(n^{\frac{1}{2}})$ states. In the case of 1-dimensional Tile Automata, we achieve the same bound with freezing, deterministic rules. 


\subsection{Freezing Deterministic Single-Transition}
We start by showing how to encode a binary string of length $n$ in a set of (freezing) transition rules that take place on a $2 \times(n + 2)$ rectangle that will print the string on its right side. We extend this construction to work for an arbitrary base string.

\subsubsection{Overview}
Consider a system that builds a length-$n$ string. First, we create a rectangle of index states that is two wide, as seen on the left side of Figure \ref{fig:indexString}.  
Each row has a unique pair of index states, so each bit of the string is uniquely indexed.
We divide the index states into two groups based on which column they are in, and which ``digit'' they represent. 
Let $r = \lceil n^\frac{1}{2}\rceil$.
Starting with index states $A_0$ and $B_0$, we build a counter pattern with base $r$. We use $\mathcal{O}(n^\frac{1}{2})$ states, shown in Figure \ref{fig:doubleStates}, to build this pattern. We encode each bit of the string in a transition rule between the two states that index that bit. A table with these transition rules can be seen in Figure \ref{tab:stringLU}.

The pattern is built in $r$ sections of size $2 \times r$ with the first section growing off of the seed. The tile in state $S_A$ is the seed. There is also a state $S_B$ that has affinity for the right side of $S_A$. 
The building process is defined in the following steps for each section.
\begin{enumerate}
	\item  The states $S_B, 0_B, 1_B,\dots , (r - 1)_B$ grow off of $S_B$, forming the right column of the section. The last $B$ state allows for $a'$ to attach on its west side. $a$ tiles attach below $a'$ and below itself. This places $a$ states in a row south toward the state $S_A$, depicted in Figure \ref{fig:detBuild}.
	\item Once a section is built, the states begin to follow their transition rules shown in Figure \ref{fig:transition3}. The $a$ state transitions with seed state $S_A$ to begin indexing the $A$ column by changing state $a$ to state $0_A$. For $1 \leq y \leq n - 2$, state $a$ vertically transitions with the other $y'_A$ states, incrementing the index by changing from state $a$ to state $(y+1)_A$.
	\item This new index state $z_A$ propagates up by transitioning the $a$ tiles to the state $z_A$ as well. Once the $z_A$ state reaches $a'$ at the top of the column, it transitions $a'$ to the state $z'_A$. Figure \ref{fig:transition4} presents this process of indexing the $A$ column. 
	\item If $z <  n - 1$, there is a horizontal transition rule from states $(z'_A, (n - 1)_B)$ to states $(z'_A, (n - 1)'_B)$. The state $0_B$ attaches to the north of $(n - 1)_B$ and starts the next section. If $z = n$, there does not exist a transition.
	\item This creates an assembly with a unique state pair in each row as seen in the first column of Figure \ref{fig:indexString}.

\end{enumerate}


\subsubsection{States}
\begin{figure}[t]
	\centering
	\includegraphics[width=1.\textwidth]{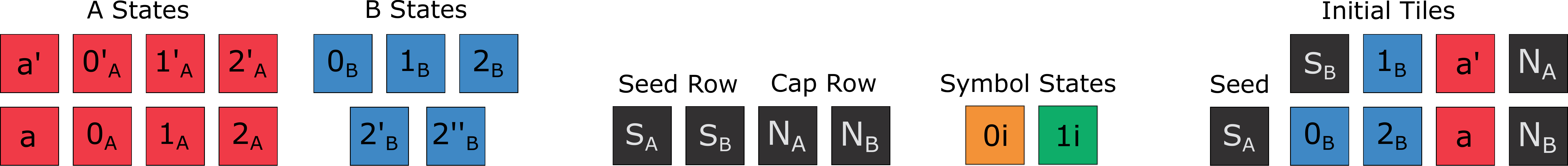}
	\caption{States to build a length-$9$ string in deterministic Tile Automata. }
	\label{fig:doubleStates}
\end{figure}

An example system with the states required to print a length-$9$ string are shown in Figure \ref{fig:doubleStates}. 
The first states build the seed row of the assembly. The seed tile has the state $S_A$ with initial tiles in state $S_B$. The index states are divided into two groups. The first set of index states, which we call the $A$ index states, are used to build the left column. For each $i$, $0 \leq i < r$, we have the states $i_A$ and $i'_A$. There are two states $a$ and $a'$, which exist as initial tiles and act as ``blank'' states that can transition to the other $A$ states. The second set of index states are the $B$ states. Again, we have $r$ $B$ states numbered from $0$ to $r-1$, however, we do not have a prime for each state. Instead, there are two states $r-1'_B$ and $r-1''_B$, that are used to control the growth of the next column and the printing of the strings. The last states are the symbol states $0_S$ and $1_S$, the states that represent the string.

\subsubsection{Affinity Rules / Placing Section}
\begin{figure}[t]
	\centering
	\begin{subfigure}[b]{.5\textwidth}
		\centering
		\includegraphics[width=0.9\textwidth]{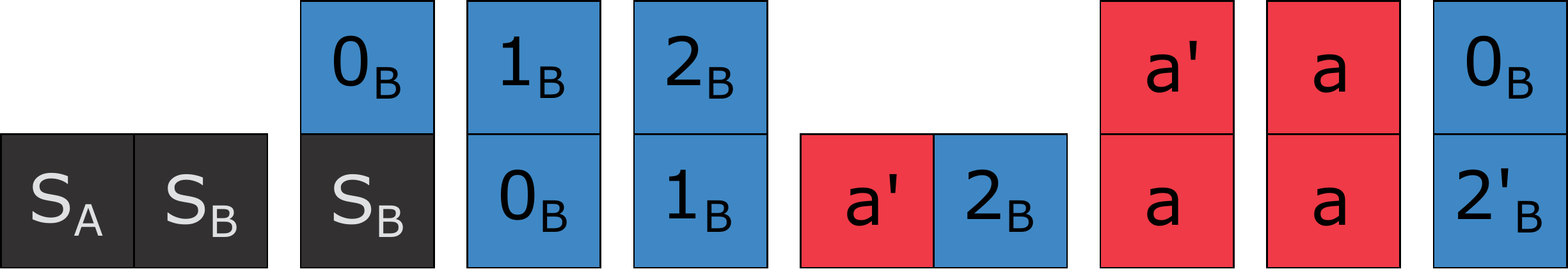}
		\caption{Affinity Rules for Initial Tiles}
		\label{fig:detAff}
	\end{subfigure}
	\begin{subfigure}[b]{.45\textwidth}
		\centering
		\includegraphics[width=0.9\textwidth]{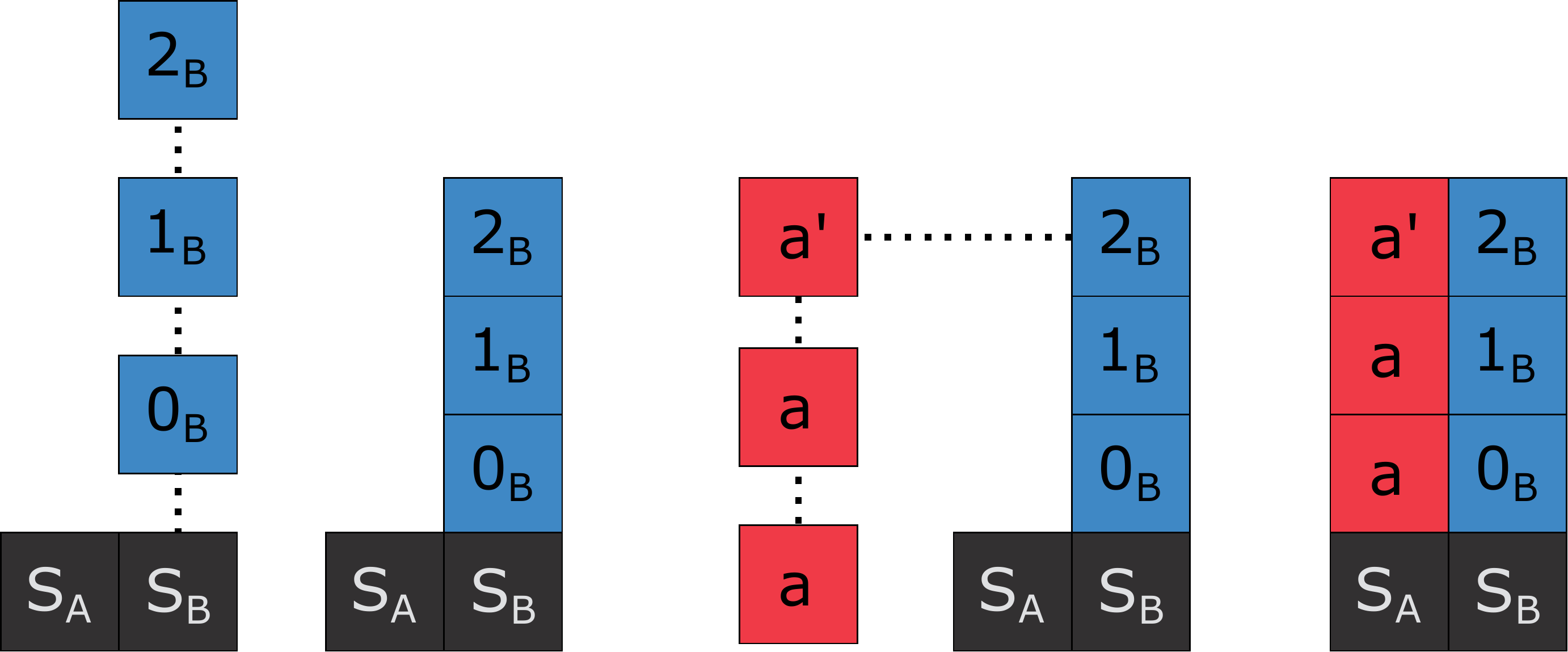}
		\caption{Process of Building a section}
		\label{fig:detBuild}
	\end{subfigure}
    \caption{(a) Affinity rules to build each section. We only show affinity rules that are actually used in the system for initial tiles to attach, but the system has more rules in order to meet the affinity-strengthening restriction. (b) The $B$ column attaches above the state $S_B$ as shown by the dotted lines. The $a'$ attaches to the left of $2_B$ and the other $a$ states may attach below it until they reach $S_A$.}
\end{figure}

Here, we describe the affinity rules for building the first section. We later describe how this is generalized to the other $r-1$ sections. We walk through this process in Figure \ref{fig:detBuild}.
To begin, the $B$ states attach in sequence above the tile $S_B$ in the seed row. Assuming $r^2 = n$, $n$ is a perfect square, the first state to attach is $0_B$. $1_B$ attaches above this tile and so on. The last $B$ state, $(r-1)_B$, does not have affinity with $0_B$, so the column stops growing. However, the state $a'$ has affinity on the left of $(r-1)_B$ and can attach. $a$ has affinity for the south side of $a'$, so it attaches below. The $a$ state also has a vertical affinity with itself. This grows the $A$ column southward toward the seed row. 

If $n$ is not a perfect square, we start the index state pattern at a different value. We do so by finding the value $q = r^2 - n$. In general, the state $i_B$ attaches above $S_B$ for $i = q \mod r$.

\subsubsection{Transition Rules / Indexing $A$ column}
\begin{figure}[t]
	\centering
	\begin{subfigure}[b]{.5\textwidth}
		\centering
		\includegraphics[width=1.\textwidth]{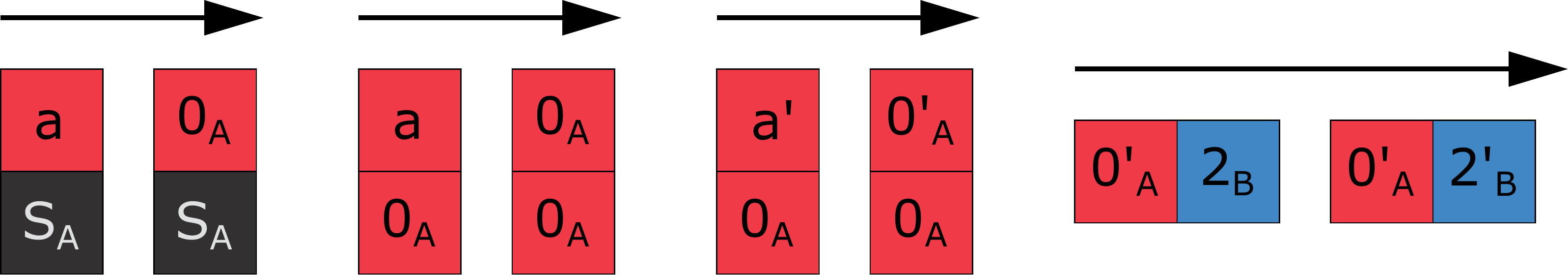}
		\caption{Transition Rules to Index the first section}
		\label{fig:transition3}
	\end{subfigure}
	\begin{subfigure}[b]{.45\textwidth}
		\centering
		\includegraphics[width=0.8\textwidth]{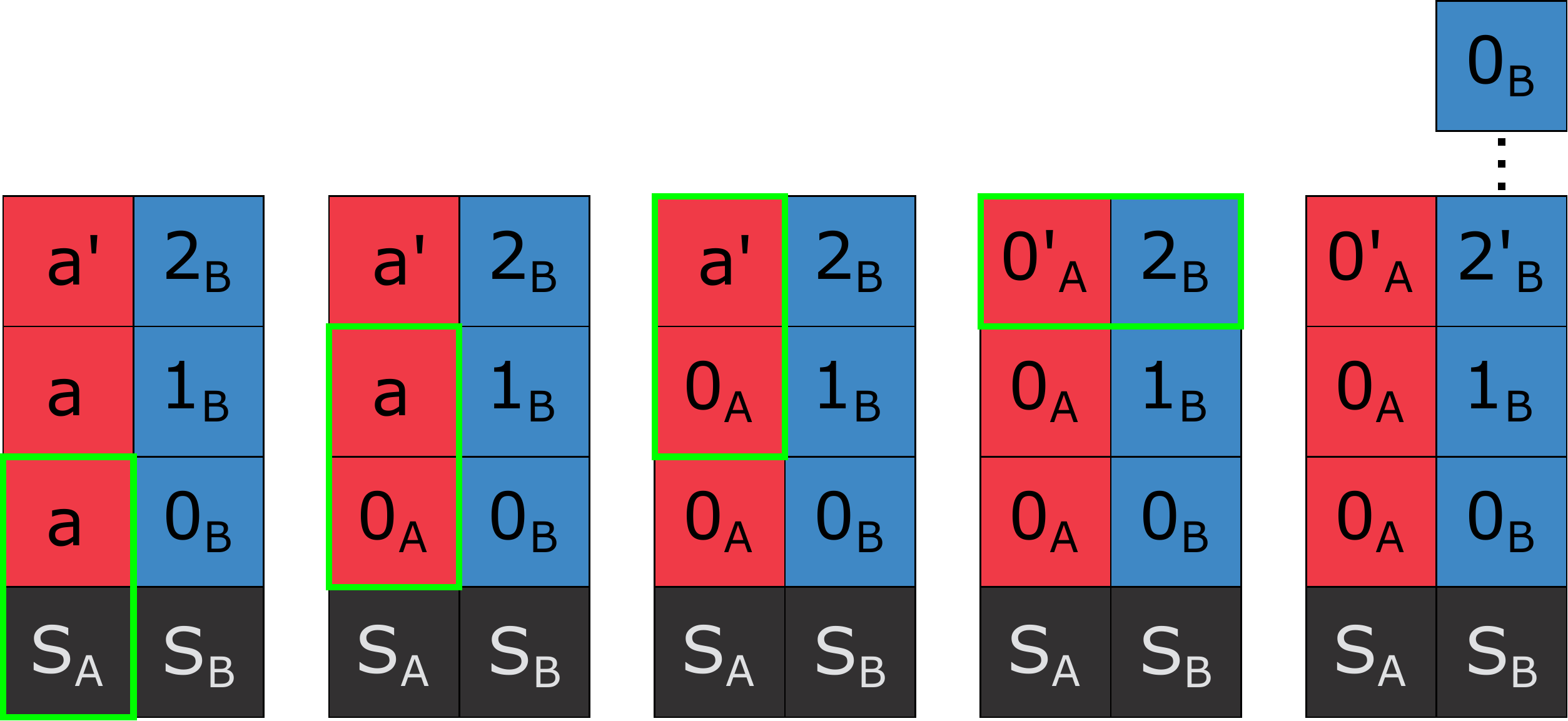}
		\caption{Process of Indexing $A$ column}
		\label{fig:transition4}
	\end{subfigure}
    \caption{(a) The first transition rule used is takes place between the seed $S_A$ and the $a$ state changing to $0_A$. The state $0_A$ changes the states north of it to $0_A$ or $0'_A$. Finally, the state $0'_A$ transitions with $2_B$ (b) Once the $a$ states reach the seed row they transition with the state $S_A$ to go to $0_A$. This state propagates upward to the top of the section.}
\end{figure}

Once the $A$ column is complete and the last $A$ state is placed above the seed, it transitions with $S_A$ to $0_A$ (assuming $r^2 = n$). $A$ has a vertical transition rule with $i_A$ ($0 \leq i < r$) changing the state $A$ to state $i_A$. This can be seen in Figure \ref{fig:transition3}, where the $0_A$ state is propagated upward to the $A'$ state. The $A'$ state also transitions when $0_A$ is below it, going from state $A'$ to state $0'_A$. If $n$ is not a perfect square, then $A$ transitions to $i_A$ for $i = \lfloor q / r \rfloor$. 

Once the transition rules have finished indexing the $A$ column if $i < r - 1$, the last state, $i'_A$, transitions with $(r-1)_B$, changing the state $(r-1)_B$ to $(r-1)'_B$. This transition can be seen in Figure \ref{fig:transition4}. The new state $(r-1)'_B$ has an affinity rule allowing $0_B$ to attach above it, and allowing the next section to be built. When the state $A$ is above a state $j'_A$,  $0 \leq j < r - 1$, it transitions with that state, thus changing from state $A$ to $(j+1)_A$,  which increments the $A$ index.

\subsubsection{Look up}
After creating a $2 \times (n + 2)$ rectangle, we encode a length-$n$ string $S$ into the transitions rules. Note that each row of our assembly consists of a unique pair of index states, which we call a \emph{bit gadget}. Each bit gadget will \emph{look up} a specific bit of our string and transition the $B$ tile to a state representing the value of that bit. 

\begin{figure}[t]
	\centering
	\begin{subfigure}[b]{.61\textwidth}
		\centering
		\includegraphics[width=1.\textwidth]{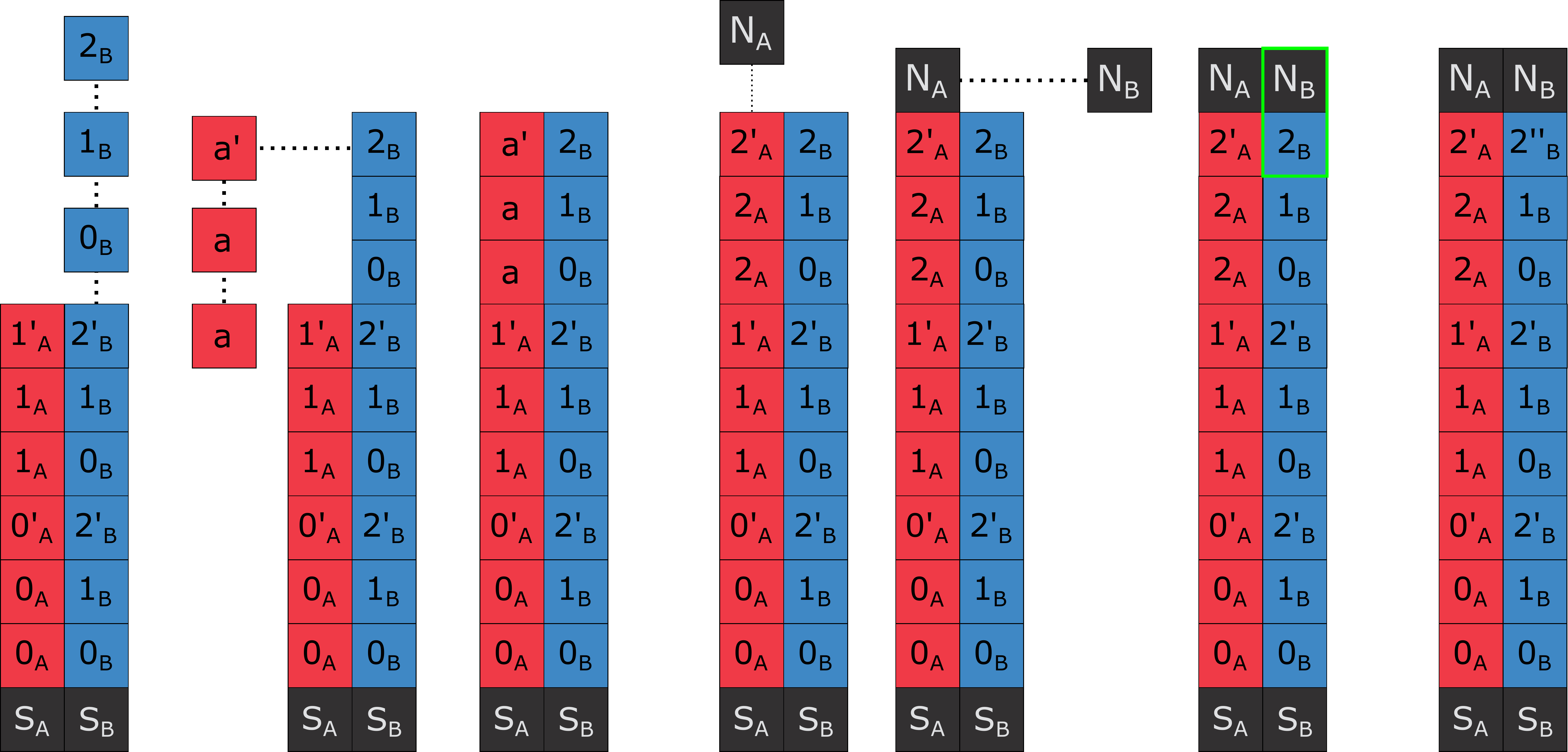}
		\caption{Attaching Cap Row}
		\label{fig:capRow}
	\end{subfigure}
	\begin{subfigure}[b]{.18\textwidth}
		\centering\footnotesize
		\begin{tabular}{c c | c}
				\textbf{A} & \textbf{B} & $S$ \\ \hline
				 2 & 2 & 0 \\
				 2 & 1 & 1 \\
				 2 & 0 & 1 \\
				 1 & 2 & 1 \\
				 1 & 1 & 0 \\
				 1 & 0 & 1 \\
				 0 & 2 & 1 \\
				 0 & 1 & 0 \\
				 0 & 0 & 0 \\
		\end{tabular}
		\caption{Encoding $S$}
		\label{tab:stringLU}
	\end{subfigure}
	\begin{subfigure}[b]{.19\textwidth}
		\centering
		\includegraphics[width=.79\textwidth]{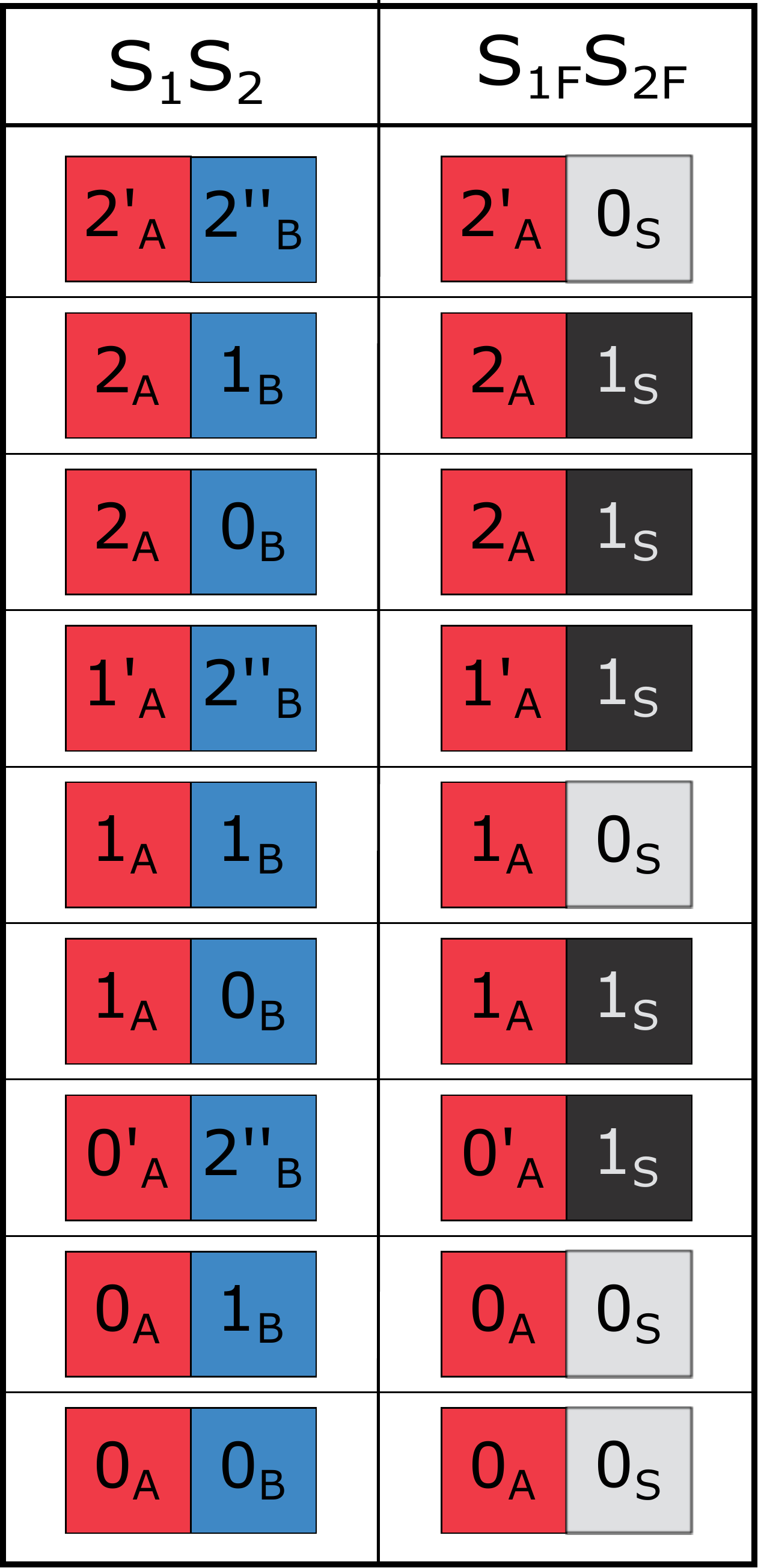}
		\caption{Transition Rules}
		\label{fig:indexString}
	\end{subfigure}
    \caption{
    (a) Once the last section finishes building the state $N_A$ attaches above $2'_A$. $N_B$ then attaches to the assembly and transitions with $2_B$ changing it directly to $2''_B$ so the string may begin printing.
    (b) A table indexing the string $S = 011101100$ using two columns and base $|S|^\frac{1}{2}$. (c) Transition Rules to print $S$. We build an assembly where each row has a unique pair of index states in ascending order.  }
\end{figure}

Figure \ref{tab:stringLU} shows how to encode a string $S$ in a table with two columns using $r$ digits to index each bit. From this encoding, we create our transition rules. Consider the $k^{th}$ bit of $S$ (where the $0^{th}$ bit is the least significant bit) for $k = ir + j$. Add transition rules between the states $i_A$ and $j_B$, changing the state $j_B$ to either $0_S$ or $1_S$ based on the $k^{th}$ bit of $S$. This transition rule is slightly different for the northmost row of each section as the state in the $A$ column is $i'_A$. Also, we do not want the state in the $B$ column, $(r-1)_B$, to prematurely transition to a symbol state. Thus, we have the two states $(r-1)'_B$ and $(r-1)''_B$. As mentioned, once the $A$ column finishes indexing, it changes the state $(r-1)_B$ to state $(r-1)'_B$, thus allowing for $0_B$ to attach above it, which starts the next column. Once the state $0_B$ (or a symbol state) is above $(r-1)'_B$, there are no longer any possible undesired attachments, so the state transitions to $(r-1)''_B$, which has the transition to the symbol state.

The last section has a slightly different process as the $(r-1)_B$ state will never have a $0_B$ attach above it, so we have a different transition rule. This alternate process is shown in Figure \ref{fig:capRow}. The state $(r-1)'_A$ has a vertical affinity with the cap state $N_A$. This state allows $N_B$ to attach on its right side. This state transitions with $(r-1)_B$ below it, changing it directly to $(r-1)''_B$, thus allowing the symbol state to print.


\begin{theorem}\label{thm:2LU}
For any binary string $S$ with length $n > 0$, there exists a freezing Tile Automata system $\Gamma_s$ with deterministic single-transition rules, that uniquely assembles a $2 \times (n + 2)$ assembly $A_S$ that represents $S$ with $\mathcal{O}(n^{\frac{1}{2}})$ states. 
\end{theorem}

\begin{proof}
Let $r = n^\frac{1}{2}$. We construct the system $\Gamma_s$ as follows. We use $\BO{r}$ index states and a constant number of seed row and symbol states as described above. The seed tile is the state $S_A$, and our initial tiles are the $B$ index states, and the two states $a$ and $a'$. The affinity rules are also described above. Due to affinity strengthening, if one state $\sigma_1$ may transition to another state $\sigma_2$, then $\sigma_1$ must have at least the same affinity rules as $\sigma_2$. This does not cause an issue for the $A$ column as the $a$ state already has affinity with itself, so all the $A$ states (besides the states at the top of the column with subscript $A'$) will have vertical affinity with each other. The top tile of each $A$ column does not have affinity with any states on its north and it never gains one by transitioning. For $0 \leq i < r - 1$,  the state $i_B$ has a vertical affinity with $(i+1)_B$. These $B$ index states transition to symbol states, so the symbol states must have vertical affinities with all the $B$ states. 

We encode the starting values of the indices in the affinity rule of $S_B$ for the $B$ index, and the transition rule between $A$ and $S_A$ for the $A$ column. In the case that the first $B$ state is $(r-1)_B$, the state that is above $S_A$ will be $a'$. In this case, we transition $a'$ with $S_A$, thus changing it directly to $0'_A$. 

Our transition rules to build the sections are described above. This system has deterministic rules since each pair of states has up to one transition rule between them. Now, we formally describe the method to encode string $S$. 

In Figure \ref{tab:stringLU}, $S$ is drawn as a table with indices of up to $r$. Let $S( \alpha, \beta  )$ be the $k^{th}$ bit of $S$ where $k = \alpha r + \beta - O$. For $\beta < r -1$, we have a transition rules between states $(\alpha_A, \beta_B)$ transitioning to $(\alpha_A, 0i)$ if $S( \alpha, \beta  ) = 0$, or to $(\alpha_A, 1i)$ if $S( \alpha, \beta  ) = 1$. When $\beta = r-1$, we have the same rule but between $(\alpha_A, \beta''_b)$. An overview of these rules can be seen in Figure \ref{fig:indexString}.

The assembly $A_S$ has the seed row states as its bottom row, followed by rows with an $A$ index state on the left and a symbol state on the right. The top row has the two cap states $N_A$ and $N_B$. This assembly is terminal since none of the initial tiles may attach to the north or south row, the $A$ index states do not have any left affinities, and the symbol states do not have any right affinities. 
This assembly is uniquely produced given how each section is built with only one available move at each step when building the first section until the $A$ column begins indexing. When the $A$ column starts indexing, it is able to change the states of the tiles in the $B$ column. The affinity strengthening requirement forces the symbol states to have affinity with all the $B$ states. This does not cause an issue for most of the $B$ states as the tiles around them have already attached, but this is why the state $(r-1)_B$ does not transition to a symbol state, yet instead waits until the next section has started (or the cap row is present) to transition to $(r-1)''_B$. 
\end{proof}

\subsubsection{Arbitrary Base}
In order to optimally build rectangles, we first print arbitrary base strings. Here, we show how to generalize Theorem \ref{thm:2LU} to print base-$b$ strings. 

\begin{corollary}\label{col:arbBase} 
For any base-$b$ string $S$ with length $n > 0$, there exists a freezing Tile Automata system $\Gamma$ with deterministic single-transition rules, that uniquely assembles an $(n + 2) \times 2$ assembly that represents $S$ with $\mathcal{O}(n^{\frac{1}{2}} + b)$ states. 
\end{corollary}
\begin{proof} 
In order to print base-$b$ strings, we use $b$ index states from $0_S$ to $(b-1)_S$. We encode the strings in our transitions the same way as the above proof by transitioning the states $(\alpha_A, \beta_B)$ to $(\alpha_A, k_S)$ for $k = S(\alpha, \beta)$.
\end{proof}

\subsubsection{Optimal Bounds}
Using Corollary \ref{col:arbBase} and base conversion techniques from \cite{adleman2001running},  we achieve the optimal bound for binary strings. The techniques from previous work extend the size of the assembly to a non-constant height. 

\begin{theorem}\label{thm:optDetStr}
For any binary string $S$ with length $n > 0$, there exists a freezing Tile Automata system $\Gamma_s$ with deterministic single-transition rules, that uniquely assembles an $n \times (n + 2)$ assembly $A_S$ that represents $S$ with $\Theta( \frac{n}{\log n}^\frac{1}{2} )$ states. 
\end{theorem}
\begin{proof}
Divide $S$ into $\frac{|S|}{\log |S|}$ segments- each of length $\log |S|$. Build a length $\frac{|S|}{\log |S|}$ base-$\frac{n}{\log n}^\frac{1}{2}$  string in $\mathcal{O}( \frac{n}{\log n}^\frac{1}{2} )$ states using Corollary \ref{col:arbBase}. Then using the base conversion technique from \cite{adleman2001running} to achieve the final binary string.  
 \end{proof}

\subsection{Height-$1$ Strings}\label{sec:thinStr}
\begin{figure}[t]
	\centering
	\begin{subfigure}[b]{.45\textwidth}
		\centering
		\includegraphics[width=0.25\textwidth]{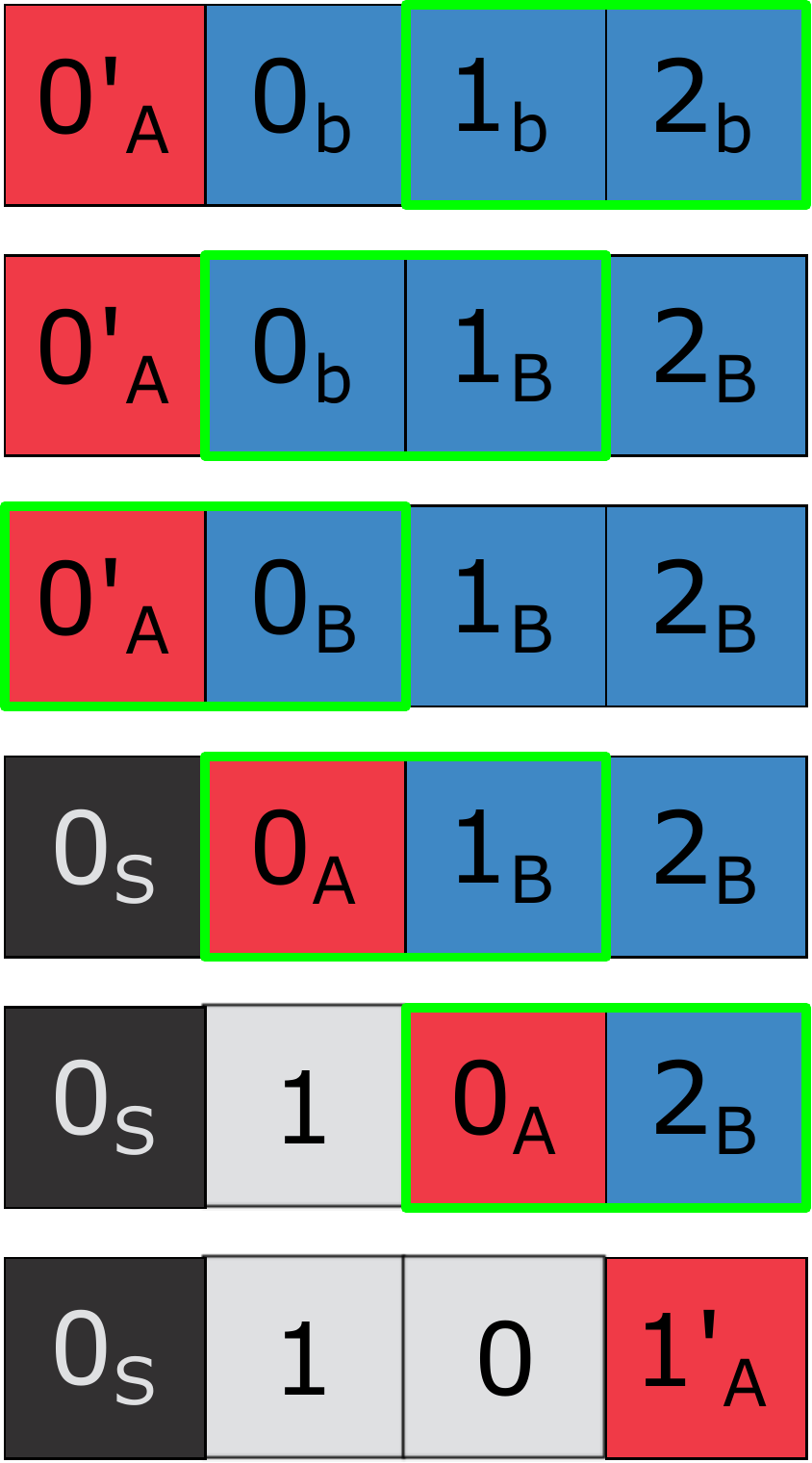}
		\caption{Transition Rules that occur on the first segment of assembly}
		\label{fig:lineSeg1}
	\end{subfigure}
	\begin{subfigure}[b]{.45\textwidth}
		\centering
		\includegraphics[width=0.75\textwidth]{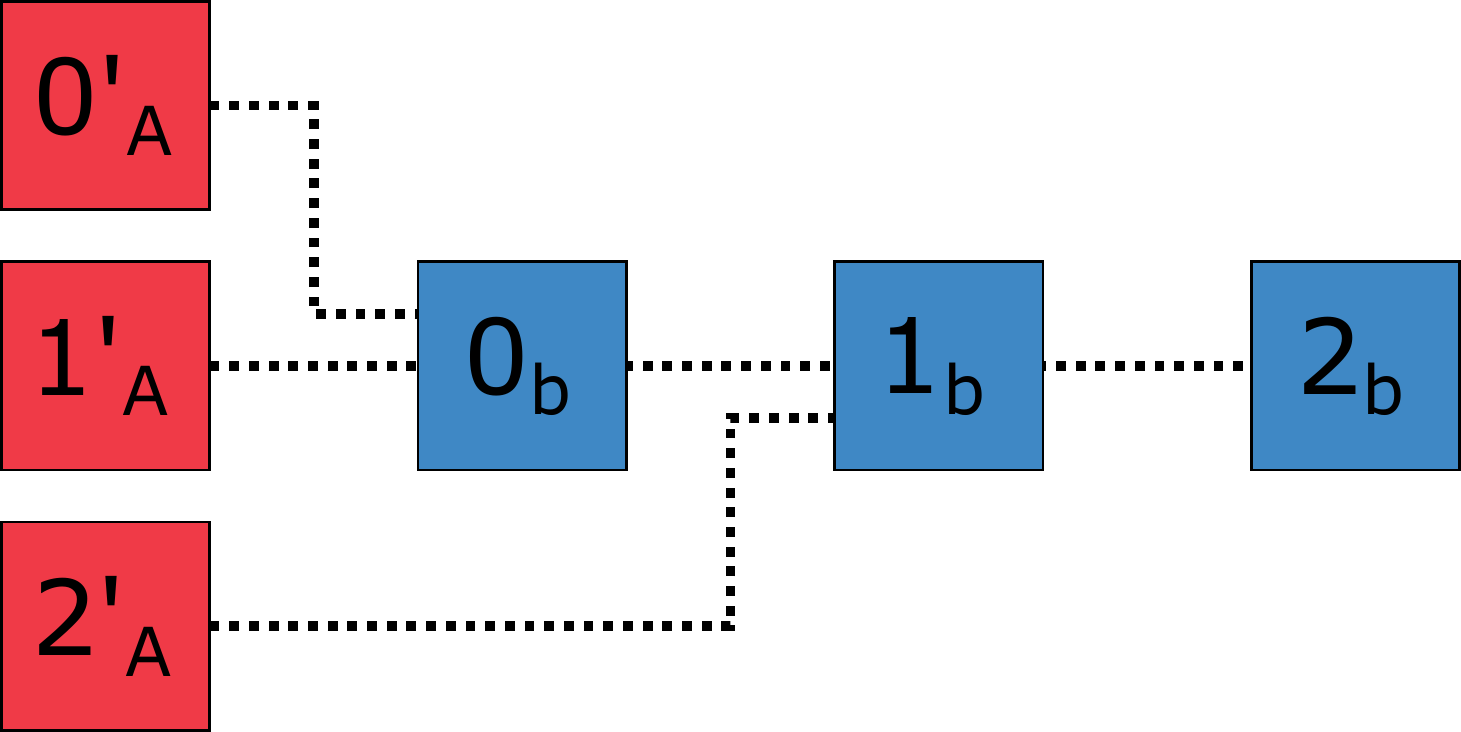}
		\caption{Affinity Rules}
		\label{fig:lineAff}
	\end{subfigure}
    \caption{(a) Once all of the $b$ tiles attach to the seed, they begin to transition to $B$ states. The $A'$ state transitions with these $B$ states to unpack the string. Once it reach the final $B$ state it increments to state $1_{A'}$ (b) Affinity rules used to build each section. The $0_b$ initial tile attach to the left of the $A'$ states, however the state $2_{A'}$ is the final segment so it has affinity with $1_b$ to achieve the exact length.  }
\end{figure}

Here, we present an alternate construction to achieve the same bound as above but at exact scale. This system, however, does not have single-transitions.
At a high level, our construction works the same way: by encoding strings in pairs of index states. However, since we are constructing an exact size $1 \times n$ assembly, we must be careful with our arrangement. We do so by building and unpacking one $1 \times r$ section at a time, which is accomplished by having a single $A$ state ``walk'' across $r$ adjacent $B$ states. Each time $A$ passes over a $B$ state, it leaves a symbol state. The process for the first section can be seen in Figure \ref{fig:lineSeg1}. 
 
\textbf{Affinity Rules.} The seed tile is $0_{A'}$, it has affinity with $0_b$ which allows the $b$ tiles to attach. Example affinity rules are shown in Figure \ref{fig:lineAff}. We include $r$ of these $b$ states to construct each full section. However since the last section does not, we have $r_{A'}$ allow $1_b$ to attach to keep the correct length. 
 
 \textbf{Transition Rules.} The states $r-1_b$ and $r_b$ transition to $r-1_B$ and $r_B$ respectively. Then each $i_b$ and $i+1_B$ transitions changing the first state to $i_B$. These transitions signal that the segment is finished building and the $A'$ state may start unpacking. The first transition that unpacks is between the seed state $0_{A'}$ and $0_B$ is a special case, since we need the left most tile to not have affinity with any other state, so this enters the $0_S$ or $1_S$ symbol state, while the $0_B$ changes to $0_A$. For $0 \leq \beta < r - 1$, the states $\alpha_A$ and $\beta_B$ transition the $\alpha_A$ to the symbol state for $S(\alpha, \beta)$ and $\beta_B$ to $\alpha_A$. If $\beta = r - 1$, then $\beta_B$ instead transitions to $\alpha+1_{A'}$. The final transition between the index states unpacks both numbers and transitions the right state to an end state marked with $T$, which does not have affinity with anything on its right, making the assembly terminal. 

\begin{figure}[t]
	\centering
	\begin{subfigure}[b]{.41\textwidth}
		\centering
		\includegraphics[width=0.6\textwidth]{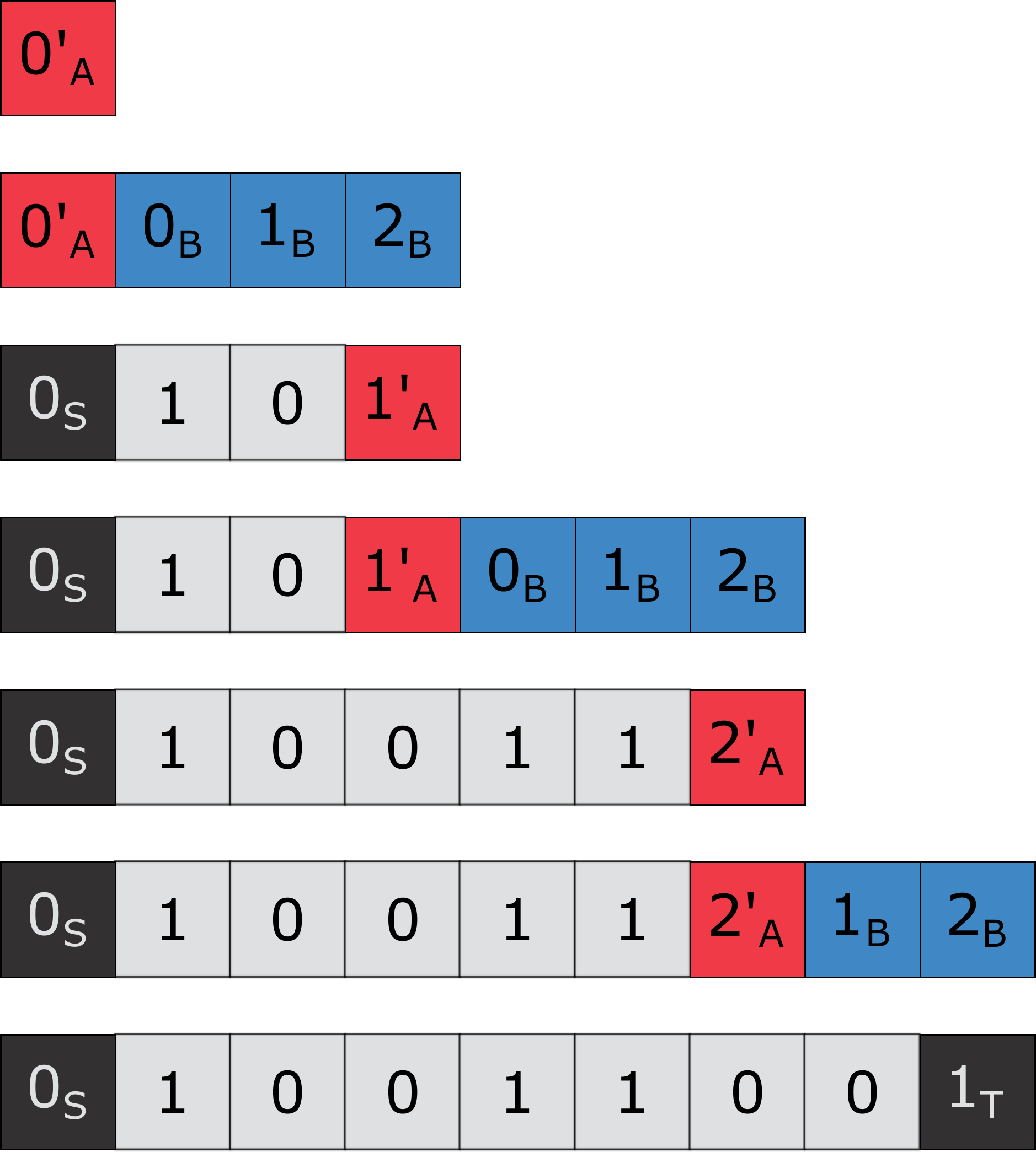}
		\caption{Construction to build string $S$}
		\label{fig:stringLine}
	\end{subfigure}
	\begin{subfigure}[b]{.53\textwidth}
		\centering
		\includegraphics[width=1.\textwidth]{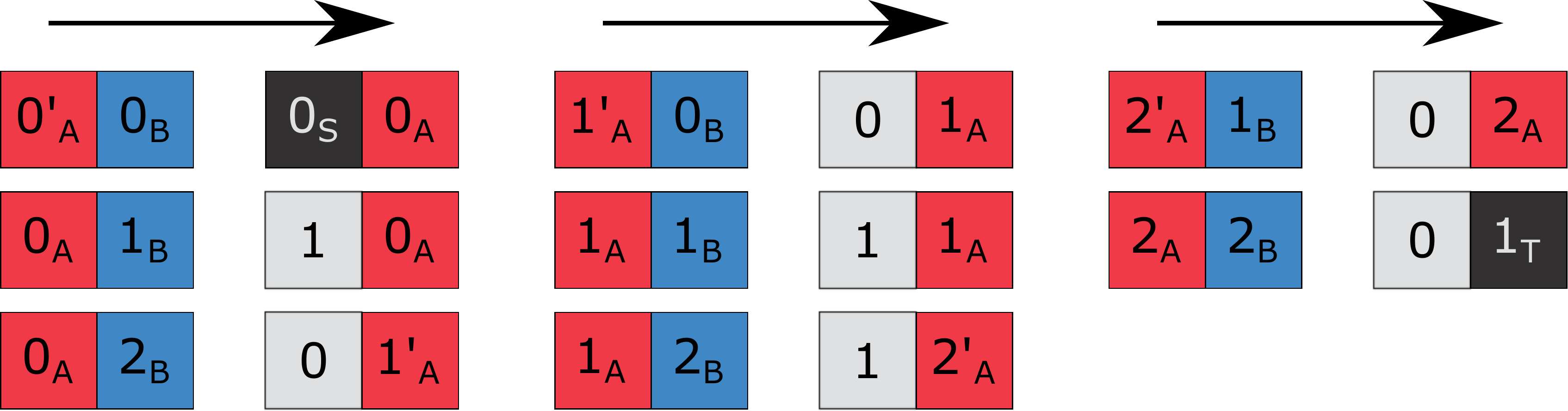}
		\caption{Transition rules to encode $S$}
		\label{fig:trLine}
	\end{subfigure}
    \caption{(a) Each segment is built using $n^\frac{1}{2}$ tiles. The $A$ state moves along the assembly leaving a symbol state in the previous location. Each time the $A$ state reaches $2_B$ it increments to an $A'$ state allowing for another segment to attach. (b) The transition rules which encode each symbol.  }
\end{figure}

\begin{theorem}\label{thm:lineString}
For any base-$b$ string $S$ with length $n > 0$, there exists a freezing Tile Automata system $\Gamma$ with deterministic transition rules, that uniquely assembles a $1 \times n$ assembly that represents $S$ with $\mathcal{O}(n^{\frac{1}{2}} + b)$ states. 
\end{theorem}
\begin{proof}
Figure \ref{fig:stringLine} outlines the process, starting from the seed tile $0_{A'}$ each section of $B$ states builds and transitions. The $A'$ state increments after reaching the final $B$ state. This allows another section to build and the process to repeat, until the last section which builds one tile shorter.  The system is freezing since each goes through the following states $b$, $B$, $A$, then finally a symbol state. The system is deterministic since each pair of states only has a single transition. While the figures depict encoding a binary string, we may use this method to encode any base symbol. 
\end{proof}
\section{Nondeterministic Transitions}\label{sec:nonDetS}


\subsection{Nondeterministic Single-Transition Systems}
For the case of single-transition systems, we use the same method from above, but instead build bit gadgets that are of size $3 \times 2$. Expanding to $3$ columns allows for a third index digit to be used, thus giving an upper bound of $\BO{n^\frac{1}{3}}$.  The second row is used for error checking, which we  describe later in the section. This system utilizes nondeterministic transitions (two states may have multiple rules with the same orientation), and is non-freezing (a tile may repeat states). This system also contains cycles in its production graph, which implies the system may run indefinitely. We conjecture this system has a polynomial run time.  
Here, let $r = \lceil n^\frac{1}{3} \rceil$.

\begin{figure}[t]
	\centering
	\begin{subfigure}[b]{.65\textwidth}
		\centering
		\includegraphics[width=1.\textwidth]{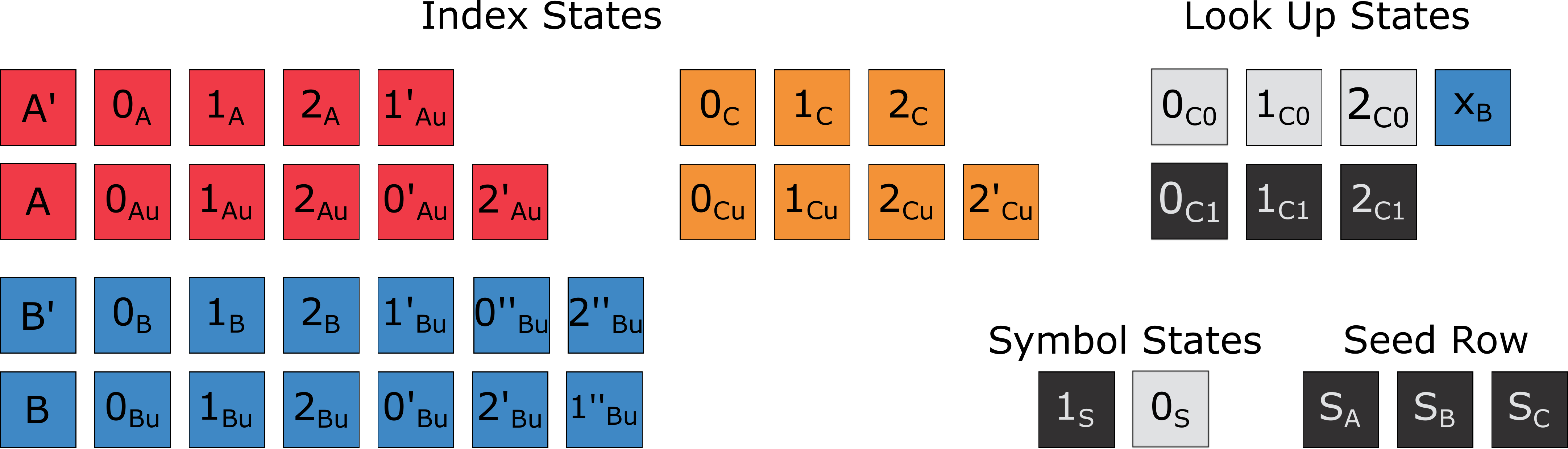}
		\caption{States space for when $|S| = 27$}
		\label{fig:stateSR}
	\end{subfigure}
	\begin{subfigure}[b]{.34\textwidth}	
	\centering
		\begin{subfigure}[b]{\textwidth}
			\centering
			\includegraphics[width=0.55\textwidth]{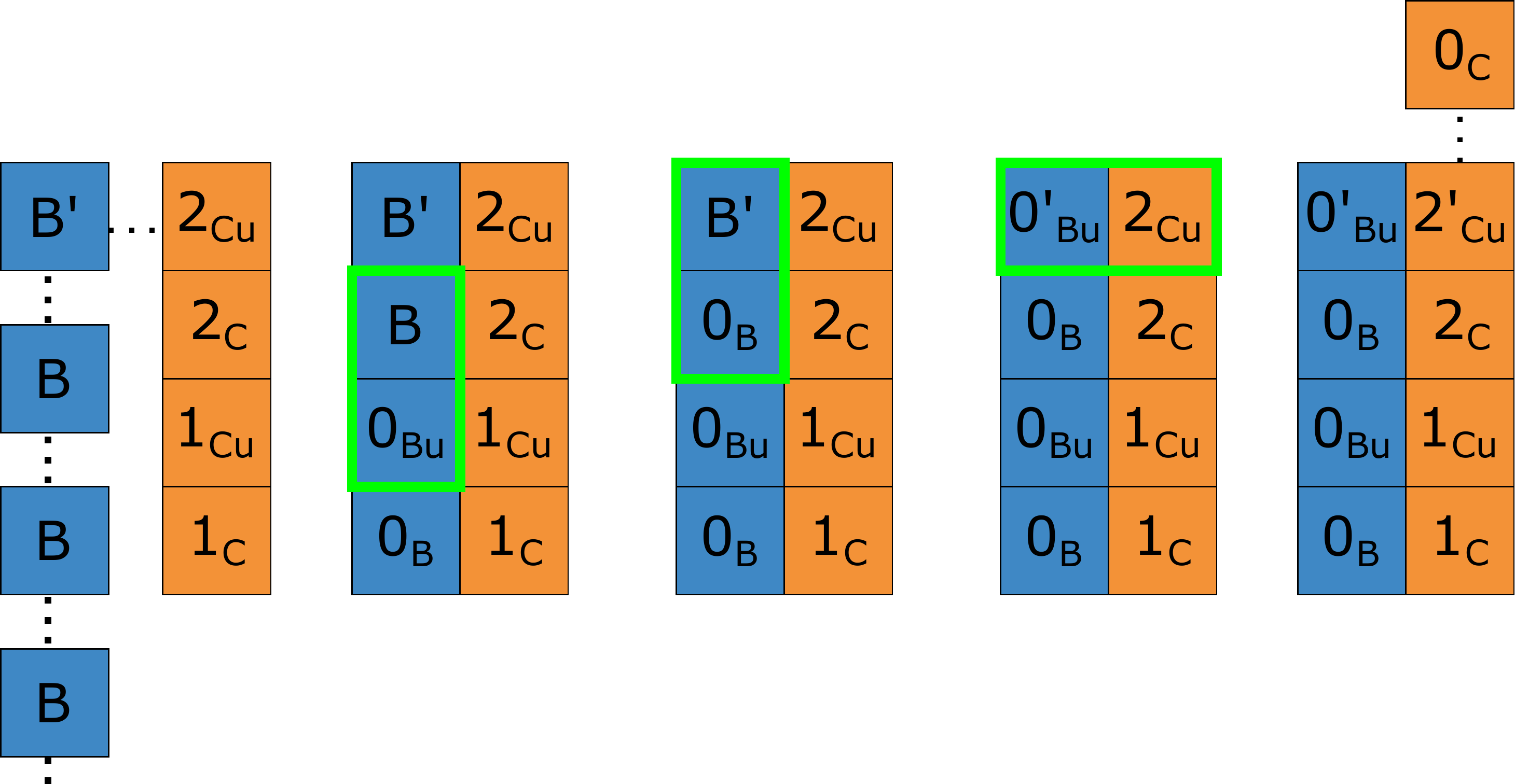}
			\caption{Indexing $B$ column}
			\label{fig:indexB}
		\end{subfigure}
		\begin{subfigure}[b]{\textwidth}
			\centering
			\includegraphics[width=.9\textwidth]{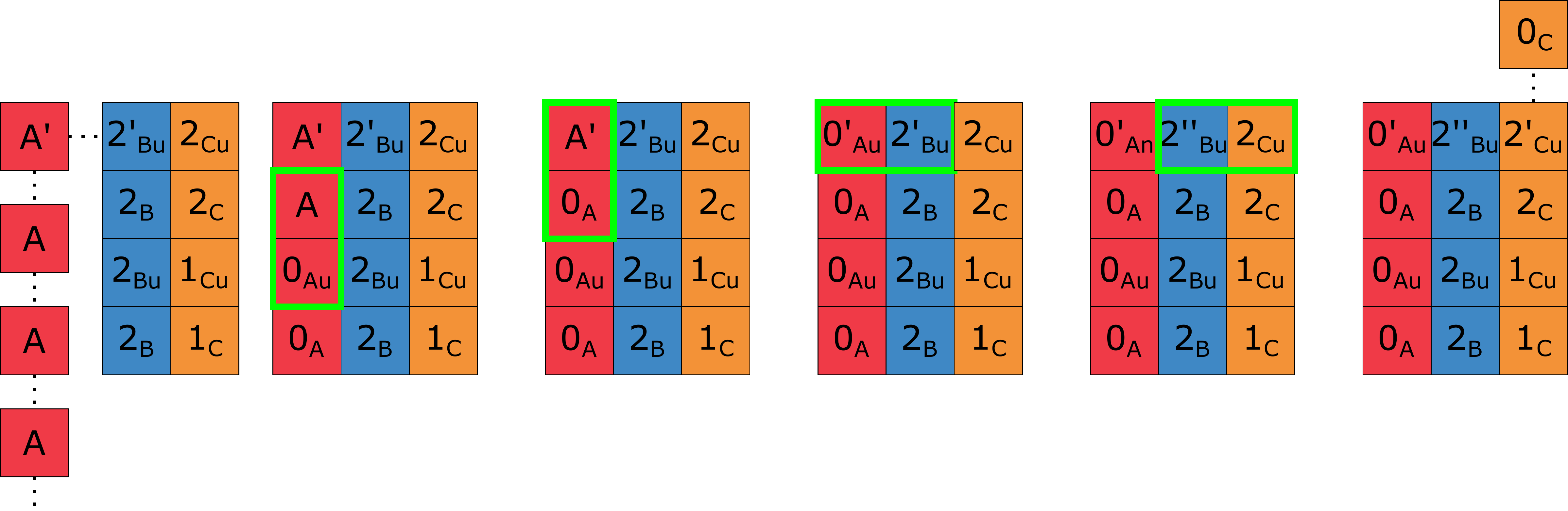}
			\caption{Indexing the $A$ column}
			\label{fig:indexA}
		\end{subfigure}
	\end{subfigure}
    \caption{(a) States needed to construct a length $27$ string where $r = 3$. (b) The index $0$ propagates upward by transitioning the tiles in the column to $0_B$ and $0_{Bu}$ and transitions $a'$ to $0'_{Bu}$. The state $0'_{Bu}$ transitions with the state $2_{Cu}$, changing the state $2_{Cu}$ to $2'_{Cu}$, which has affinity with $0_C$ to build the next section. These rules also exist for the index $1$. 
    (c) When the index state $2_B$ reaches the top of the section, it transitions $b'$ to $2'_{Bu}$. This state does not transition with the $C$ column and instead has affinity with the state $a'$, which builds the $A$ column downward. The index propagates up the $A$ column in the same way as the $B$ column. When the index state $0_A$ reaches the top of the section, it transitions the state $2'_B$ to $2''_B$. This state transitions with $2_{Cu}$ changing it to $2'_{Cu}$ allowing the column to grow. }
\end{figure}

%

\subsubsection{Index States and Look Up States}

We generalize the method from above to start from a $C$ column. The $B$ column now behaves as the second index of the pattern and is built using $B'$ and $B$ as the $A$ column was in the previous system. Once the $B$ reaches the seed row, it is indexed with its starting value. This construction also requires bit gadgets of height $2$, so we use index states $i_A, i_B, i_C$ and north index states $i_{Au}, i_{Bu}, i_{Cu}$ for $0 \leq i < r$. This allows us to separate the two functions of the bit gadget into each row. The north row has transition rules to control the building of each section. The bottom row has transition rules that encode the represented bit. 

In addition to the index states, we use $2r$ look up states, $0_{Ci}$ and $1_{Ci}$ for $0 \leq i < r$. These states are used as intermediate states during the look up. The first number ($0$ or $1$) represents the value of the retrieved bit, while the second number represents the $C$ index of the bit. The $A$ and $B$ indices of the bit will be represented by the other states in the transition rule. 

In the same way as the previous construction, we build the rightmost column first. We include the $C$ index states as initial states and allow $0_C$ to attach above $S_C$. We include affinity rules to build the column northwards as follows starting with the southmost state $0_C, 0_{Cu}, 1_C, 1_{Cu}, \dots ,r-2_{Cu}, r-1_C, r-1_{Cu}$ . 

To build the other columns, the state $b'$ can attach on the left of $r-1_{Cu}$. The state $b$ is an initial state and attaches below $b'$ and itself to grow downward toward the seed row. The state $b$ transitions with the seed row as in the previous construction to start the column. However, we alternate between $C$ states and $Cu$ states. The state $b$ above $i_C$ transitions $b$ to $i_{Cu}$. If $b$ is above $i_{Cu}$ it transitions to $i_C$. The state $b'$ above state $i_B$ transitions to $i'_{Bu}$. 
If $i < r -1$, the state $i'_B$ and $r-1_{Cu}$ transition horizontally changing $r-1'_{Cu}$, which allows $0_C$ to attach above it to repeat the process. This is shown in Figure \ref{fig:indexB}.

The state $a'$ attaches on the left of $r-1_{Cu}$. The $A$ column is indexed just like the $B$ column. For $0 \leq i < r - 1$, the state $i'_{Au}$ and $r-1'_{Bu}$ change the state $r-1'_{Bu}$ to $r-1''_{Bu}$. This state transitions with $r-1_{Cu}$, changing it to $r-1'_{Cu}$. See Figure \ref{fig:indexA}.


\subsubsection{Bit Gadget Look Up}
\begin{figure}[t]
	\centering
	\begin{subfigure}[b]{.25\textwidth}
	    \centering
        \includegraphics[width=1.\textwidth]{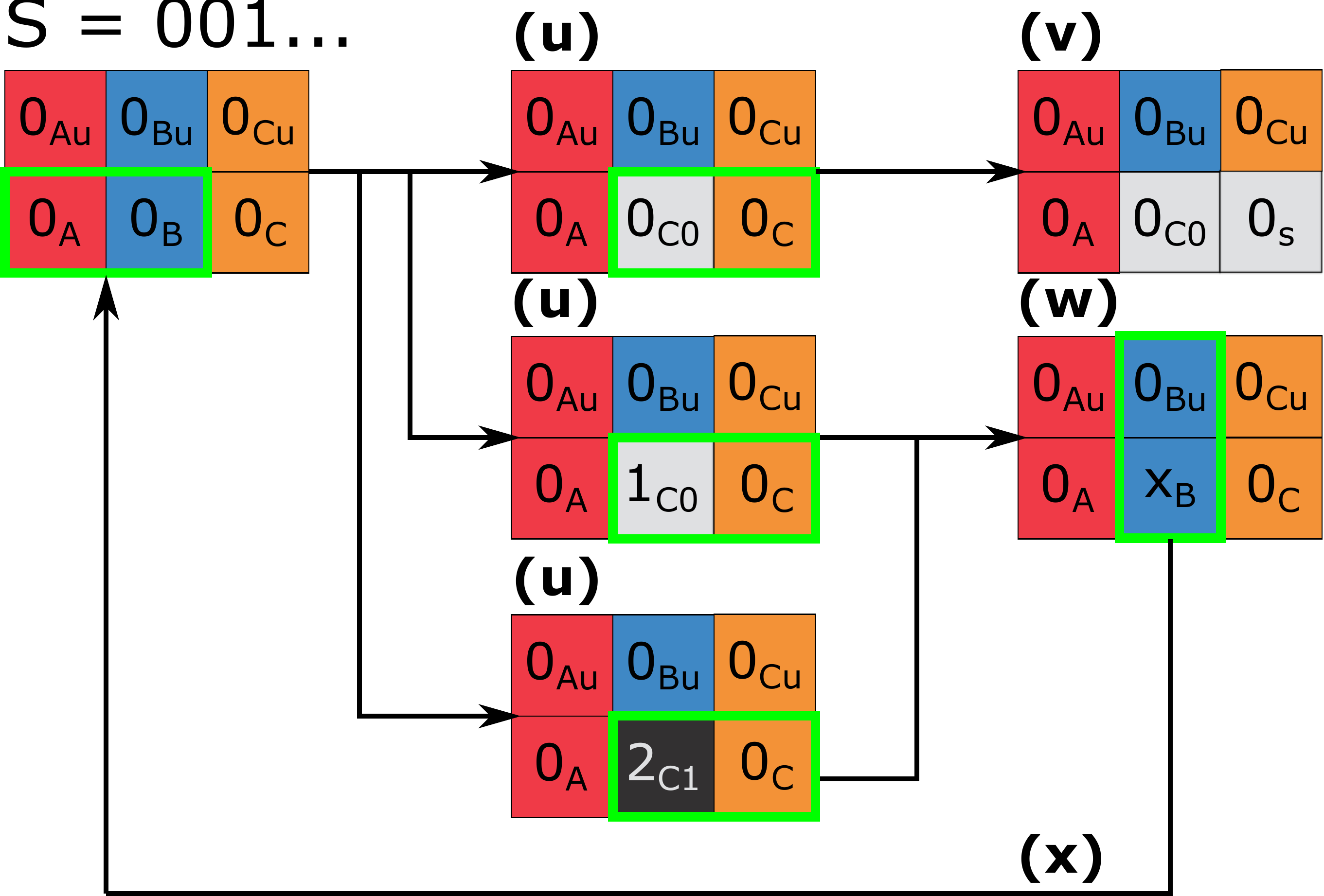}
        \caption{ST Bit Gadget look up}
        \label{fig:singleRuleBG}
    \end{subfigure}
    \begin{subfigure}[b]{.7\textwidth}
        \centering
        \includegraphics[width=.95\textwidth]{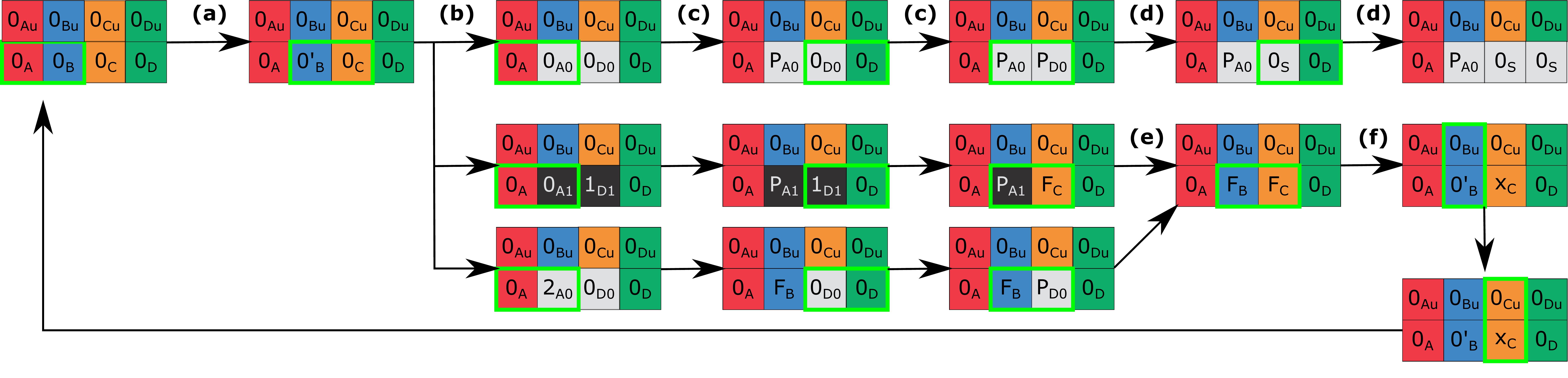}
        \caption{Nondeterministic Bit Gadget look up}
        \label{fig:flowND}
    \end{subfigure}
    \caption{(a.u) For a string $S$, where the first $3$ bits are $001$, the states $0_A$ and $0_B$ have $|S|^\frac{1}{3}$ transition rules changing the state $0_B$ to a state representing one of the first $|S|^\frac{1}{3}$ bits. The state is $i_{C0}$ if the $i^{th}$ bit is $0$ or $i_{C1}$ if the $i^{th}$ bit is $1$ (a.v) The state $0_{C0}$ and the state $0_C$ both represent the same $C$ index so the $0_C$ state transition to the $0_s$. (a.w) For all states not matching the index of $0_C$, they transition to  $x_B$, which can be seen as a blank $B$ state. (a.x) The state $0_{Bu}$ transitions with the state $x_B$ changing to $0_B$ resetting the bit gadget. 
    (b.a) Once the state $A_0$ appears in the bit gadget it transitions with $0_B$ changing $0_B$ to $0'_B$. (b.b) The states $0'_B$ and $0_C$ nondeterministically look up bits with matching $B$ and $C$ indices. The state $0'_B$ transitions to look up state representing the bit retrieved and the bit's $A$ index. The state $0_C$ transitions to a look up state representing the $D$ index of the retrieved bit. (b.c)  The look-up states transition with the states $0_A$ and $0_D$, respectively. As with the single-transition construction these may pass or fail. (b.d) When both tests pass, they transition the $D$ look up state to a symbol state that propagates out. (b.e) If a test fails, the states both go to blank states. (b.f) The blank states then reset using the states to their north.  
    }
\end{figure}

The bottom row of each bit gadget has a unique sequence of states, again we use these index states to represent the bit indexed by the digits of the states. However, since we can only transition between two tiles at a time, we must read all three states in multiple steps. These steps are outlined in Figure \ref{fig:singleRuleBG}. 
The first transition takes place between the states $i_A$ and $j_B$. We refer to these transition rules as look up rules. We have $r$ look up rules between these states for $0 \leq k < r$  of these states that changes the state $j_B$ to that state $k_{C0}$ if the bit indexed by $i, j, $ and $k$ is $0$ or the state $k_{C1}$ if the bit is $1$. 

Our bit gadget has nondeterministically looked up each bit indexed by it's $A$ and $B$ states. Now, we must compare the bit we just retrieved to the $C$ index via the state in the $C$ column. The states $k_{C0}$ and $k_C$ transition changing the state $k_C$ to the $0i$ state only when they represent the same $k$. The same is true for the state $k_{C1}$ except $C_k$ transitions to $1i$. 

If they both represent different $k$, then the state $k_C$ goes to the state $B_x$. This is the error checking of our system. The $B_x$ states transitions with the north state $j_{Bu}$ above it transitioning $B_x$ to $j_B$ once again. This takes the bit gadget back to it's starting configuration and another look up can occur. 

\begin{theorem}\label{thm:SRstrings}
For any binary string $S$ with length $n > 0$, there exists a single-transition Tile Automata system $\Gamma$, that uniquely assembles an $(2n + 2) \times 3$ assembly which represents $S$ with $\BO{n^\frac{1}{3}}$ states.  
\end{theorem}

\begin{proof} 
Let $r = n^\frac{1}{3}$, note that we use $\BO{r}$ index states and look up states in our system. The number of other states is our system is bounded by a constant so the total number of states in $\Gamma$ is  $\BO{n^\frac{1}{3}}$. We use affinity and transition rules to place tiles and index the columns as described above. 

Note that the transition rules to index column and the transition rules to signal for another section to build only ever change one of the states involved. The transition rules for the look up do the same as well. Only the $B$ state changes to a look up state. When the look up state transitions with $C$, if they both represent the same index, $C$ transitions to the symbol state of the retrieved bit. If they have different indices, then the look up state transitions to $Bx$. $Bx$ transitions with the index state above it resetting itself. All of these rules only ever change one tile.

As with Theorem \ref{thm:2LU}, the northmost and southmost rows do not allow other states to attach above/below them, respectively. The states in the $A$ column do not allow tiles to attach on their left. The rightmost column has only $Cu$ and symbol states ($1s$ or $0s$), which do not allow tiles to attach on their right. 

The last column to be indexed in this construction is the $A$ column. Bit gadgets also do not begin to transition until the $A$ state is indexed, so no matter the build order the states that are used will be present before the gadget begins transitioning. Until the $A$ column begins indexing, there is only one step that can take place so we know there are not build orders that result in other terminal assemblies. 
\end{proof}



\subsection{General Nondeterministic Transitions}
Using a similar method to the previous sections, we build length $n$ strings using $\BO{n^{\frac{1}{4}}}$ states. We start by building a pattern of index states with bit gadgets of height $2$ and width $4$. 

\subsubsection{Overview}
Here, let $r =  \lceil n^\frac{1}{4}  \rceil$. We build index states in the same way as the single-transition system but instead starting from the $D$ column. We have $4$ sets of index states, $A$, $B$, $C$, $D$. The same methods are used to control when the next section builds by transitioning the state $r-1_D$ to $r-1'_D$ when the current section is finished building.  


We use a similar look up method as the previous construction where we nondeterministically retrieve a bit. However, since we are not restricting our rules to be a single-transition system, we may retrieve $2$ indices in a single step. We include $2$ sets of $\BO{r}$ look up states, the $A$ look up states and the $D$ look up states. We also include Pass and Fail states $F_B, F_C, P_{A0}, P_{D0}, P_{A1}, P_{D1}$ along with the blank states $B_x$ and $C_x$. We utilize the same method to build the north and south row. 

Let $S(\alpha, \beta, \gamma, \delta)$ be the $i^{th}$ bit of $S$ where  $i = \alpha r^3 +  \beta r^2 + \gamma r + \delta$. The states $\beta'_B$ and $\gamma_C$ have $r^2$ transitions rules. The process of these transitions is outlined in Figure \ref{fig:flowND}. They transition from $(\beta'_B, \gamma_C)$ to either $(\alpha_{A0}, \delta_{D0})$ if $S(\alpha, \beta, \gamma, \delta) = 0$, or $(\alpha_{A1}, \delta_{D1})$ if $S(\alpha, \beta, \gamma, \delta) = 1$. After both transitions have happened, we test if the indices match to the actual $A$ and $D$ indices. We include the transition rules $(\alpha_A, \alpha_{A0})$ to $(\alpha_A, P_{A0})$ and  $(\alpha_A, \alpha_{A1})$ to $(\alpha_A, P_{A1})$. We refer to this as the bit gadget passing a test. The two states $(P_{A0}, P_{D0})$ horizontally transition to $(P_{A0}, 0s)$. The $0s$ state then transitions the state $\delta_D$ to $0s$ as well as propagating the state to the right side of the assembly. If the compared indices are not equal, then the test fails and the look up states will transition to the fail states $F_B$ or $F_C$. These fail states will transition with the states above them, resetting the bit gadget as in the previous system.

\begin{theorem}\label{thm:nonDetstrings}
For any binary string $S$ with length $n > 0$, there exists a Tile Automata system $\Gamma$, that uniquely assembles an $(2n + 2) \times 4$ assembly which represents $S$ with $\mathcal{O}(n^{\frac{1}{4}} )$ states. 
\end{theorem}

\begin{proof}
Let $r = \lceil n^\frac{1}{4} \rceil$. We use $\BO{r}$ index states to build our bit gadgets. We have $\BO{r}$ look up states and a constant number of pass, fail, and blank states to perform the look up. 

This system uniquely constructs a $(2n + 2) \times 4$ rectangle that represents the string $S$, each bit gadget is a constant height and represents a unique bit of the string. In each bit gadget the test ensure that only the correct bit is retrieved and propagated to the right side of the assembly. By the same argument as the previous constructions this assembly is terminal. A bit gadget does not begin looking up bits until its $A$ column is complete (by transitioning $\beta_B$ to $\beta'_B$) so the system uniquely constructs this rectangle as one move can happen at a time to build the bit gadgets and once a bit gadget is complete it does not affect surrounding tiles as transitions only occur between tiles in the same gadget. 
\end{proof}
\section{Rectangles}\label{sec:rectangles}
In this section, we show how to use the previous constructions to build $\mathcal{O}(\log n) \times n$ rectangles. All of these constructions rely on using the previous results to encode and print a string then adding additional states and rules to build a counter.

\begin{figure}[t]
	\centering
	\begin{subfigure}[b]{.47\textwidth}
		\centering
		\includegraphics[width=1.\textwidth]{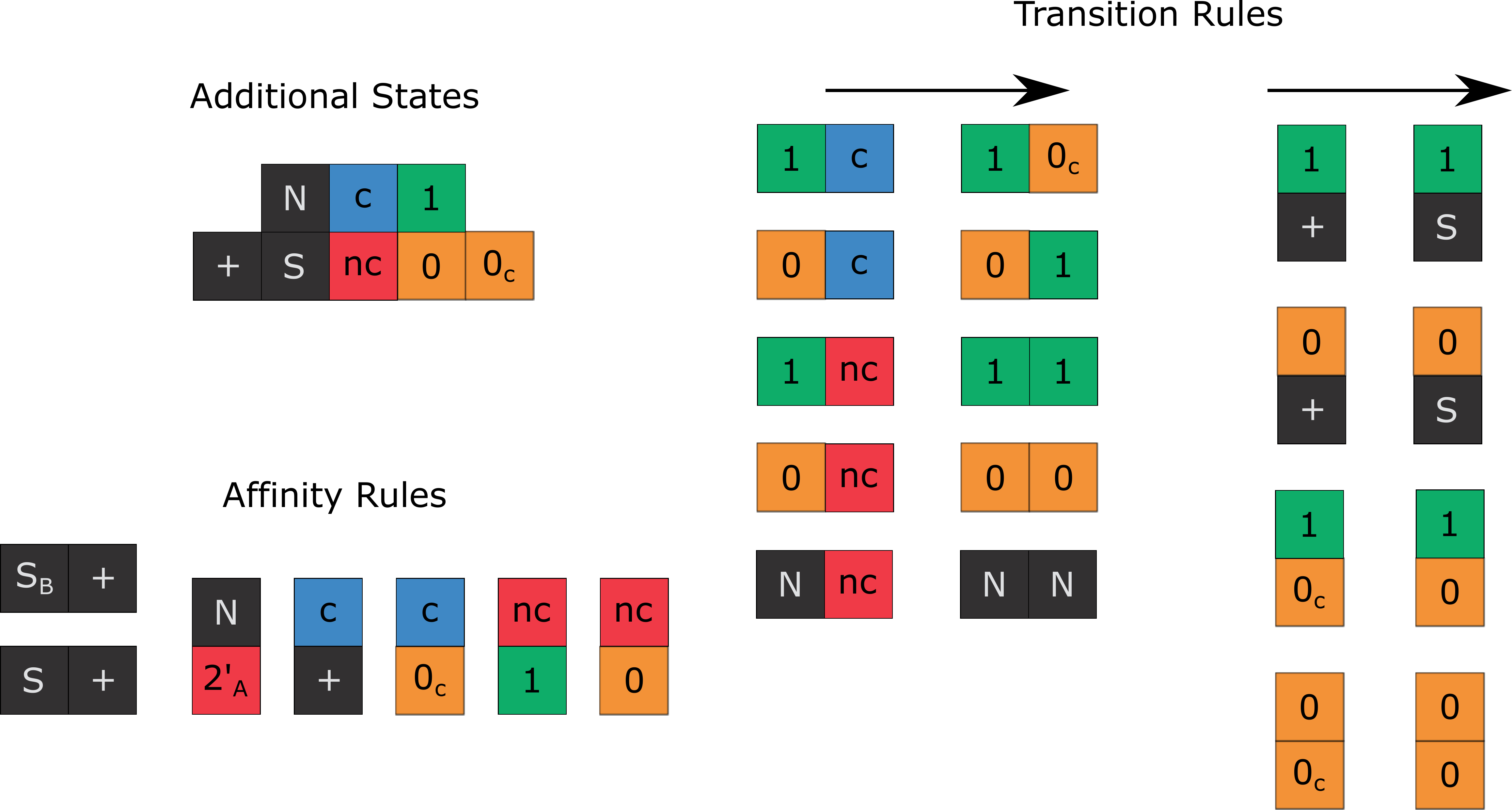}
		\caption{New states and rules for a binary counter}
		\label{fig:binCountStates}
	\end{subfigure}
	\begin{subfigure}[b]{.47\textwidth}
		\centering
		\includegraphics[width=1.\textwidth]{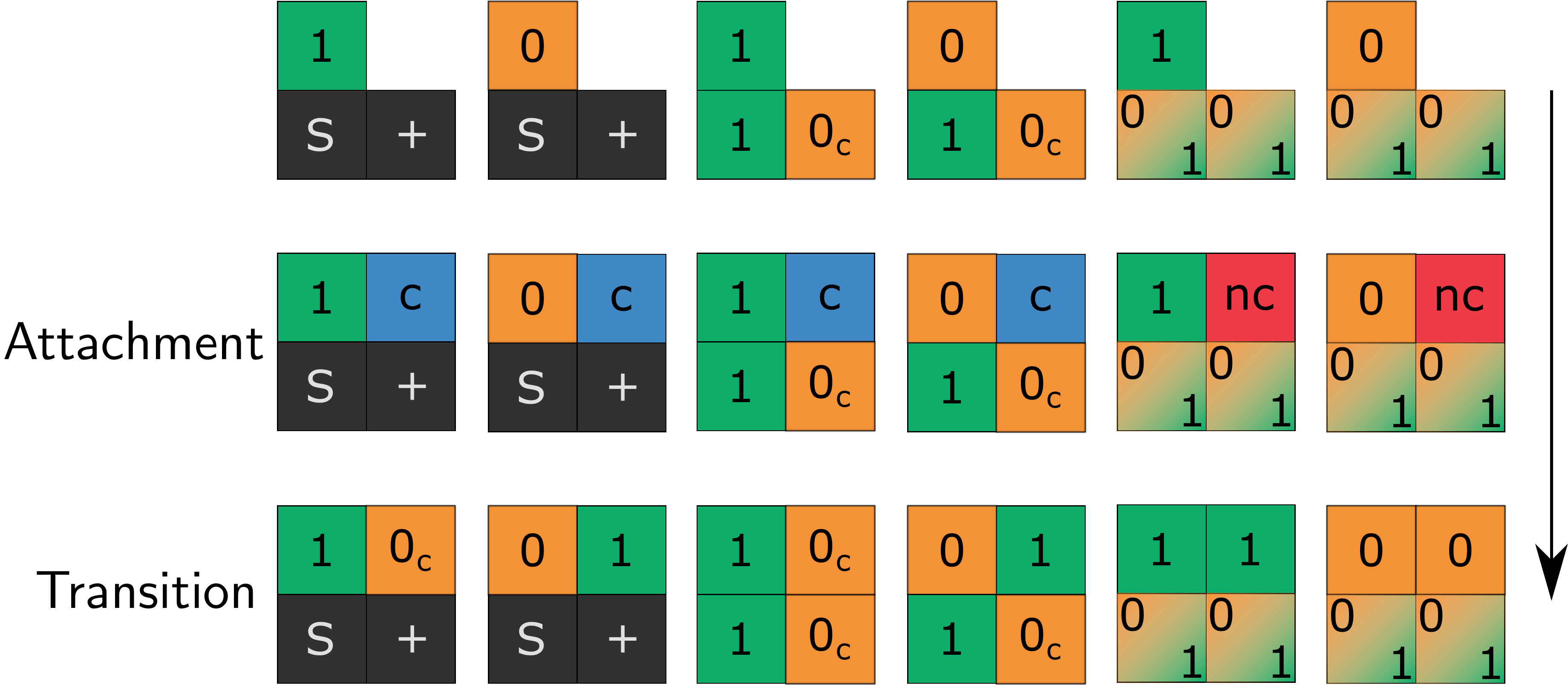}
		\caption{Every case for the half adder.}
		\label{fig:binCountAdder}
	\end{subfigure}
	\caption{(b) The 0/1 tile is not present in the system. It is used in the diagram to show that either a 0 tile or a 1 tile can take that place.}
\end{figure}

\begin{figure}[t]
    \centering
    \begin{subfigure}[b]{.45\textwidth}
        \centering
        \includegraphics[width=0.9\textwidth]{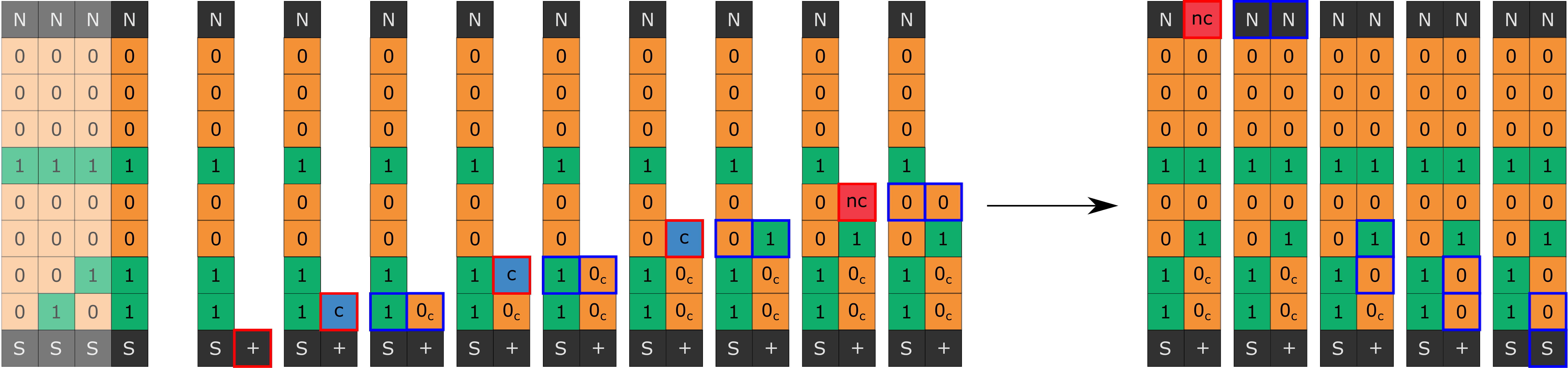}
        \caption{Binary Counter}
        \label{fig:binCountExample}
	\end{subfigure}
	\begin{subfigure}[b]{.5\textwidth}
        \centering
        \includegraphics[width=1.\textwidth]{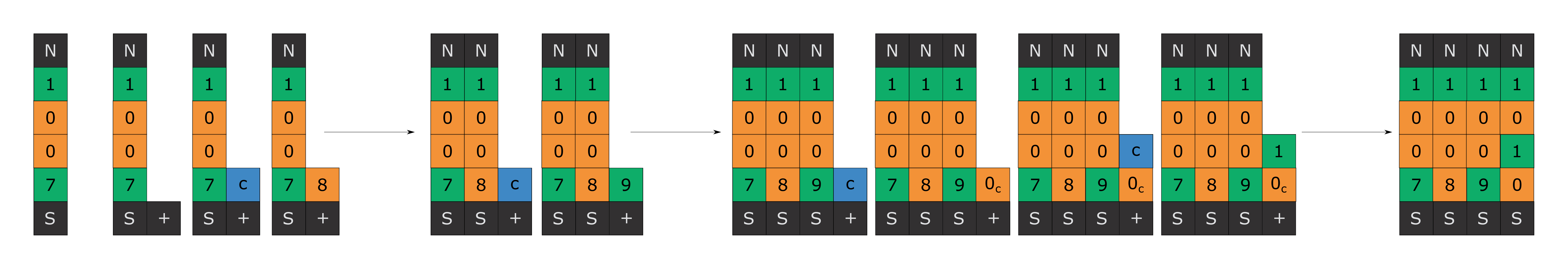}
        \caption{Base-10 Counter}
        \label{fig:decCountExample}
	\end{subfigure}
	\caption{(a) The process of the binary counter. (b) A base-10 counter.}
\end{figure}

\subsection{States}
We choose a string and construct a system that will create that string, using the techniques shown in the previous section. We then add states to implement a binary counter that will count up from the initial string. The states of the system, seen in Figure \ref{fig:binCountStates}, have two purposes. The north and south states (N and S) are the bounds of the assembly. The plus, carry, and no carry states ($+$, c, and nc) forward the counting. The 1, 0, and 0 with a carry state make up the number. The counting states and the number states work together as half adders to compute bits of the number.

\subsection{Transition Rules / Single-Tile Half-Adder }
As the column grows, in order to complete computing the number, each new tile attached in the current column along with its west neighbor are used in a half adder configuration to compute the next bit. Figure \ref{fig:binCountAdder} shows the various cases for this half adder.

When a bit is going to be computed, the first step is an attachment of a carry tile or a no-carry tile (c or nc). A carry tile is attached if the previous bit has a carry, indicated by a tile with a state of plus or 0 with a carry ($+$ or 0c). A no-carry tile is placed if the previous bit has no-carry, indicated by a tile with a state of 0 or 1. Next, a transition needs to occur between the newly attached tile and its neighbor to the west. This transition step is the addition between the newly placed tile and the west neighbor. The neighbor does not change states, but the newly placed tile changes into a number state, 0 or 1, that either contains a carry or does not. This transition step completes the half adder cycle, and the next bit is ready to be computed.

\subsection{Walls and Stopping}
The computation of a column is complete when a no-carry tile is placed next to any tile with a north state. The transition rule changes the no-carry tile into a north state, preventing the column from growing any higher. The tiles in the column with a carry transition to remove the carry information, as it is no longer needed for computation. A tile with a carry changes states into a state without the carry. The next column can begin computation when the plus tile transitions into a south tile, thus allowing a new plus tile to be attached. The assembly stops growing to the right when the last column gets stuck in an unfinished state. This column, the stopping column, has carry information in every tile that is unable to transition. When a carry tile is placed next to a north tile, there is no transition rule to change the state of the carry tile, thus preventing any more growth to the right of the column.

\begin{theorem}\label{thm:DetRect}
For all $n > 0$, there exists a Tile Automata system that uniquely assembles a $\mathcal{O}(\log{n}) \times n$ rectangle using, 

\begin{itemize}
	\item Deterministic Transition Rules and $\mathcal{O}(\log^{\frac{1}{2}}{n})$ states.
	\item Single-Transition Rules and $\Theta(\log^{\frac{1}{3}}{n})$ states.
	\item Nondeterministic Transition Rules and $\Theta(\log^{\frac{1}{4}}{n})$ states.
\end{itemize}
\end{theorem}

%
%

\subsection{Arbitrary Bases}
Here, we generalize the binary counter process for arbitrary bases. The basic functionality remains the same. The digits of the number are computed one at a time going up the column. If a digit has a carry, then a carry tile attaches to the north, just like the binary counter. If a digit has no carry, then a no-carry tile is attached to the north. The half adder addition step still adds the newly placed carry or no-carry tile with the west neighbor to compute the next digit. This requires adding $\BO{b}$ counter states to the system, where $b$ is the base.


\begin{theorem}\label{thm:varBaseDetRec}
For all $n > 0$, there exists a deterministic Tile Automata system that uniquely assembles a $\mathcal{O}(\frac{\log n}{\log \log n}) \times n$ rectangle using  $\Theta\left((\frac{\log n}{\log \log n})^\frac{1}{2}\right)$ states.
\end{theorem}

\begin{proof}
Note that all logarithms shown are base $2$. There are two cases. Case 1: $n \leq 2^{16}$. Let $c$ be the minimum number of states such that for all $w < n' < 2^{16}$, a $w \times n'$ rectangle can be uniquely assembled with $\leq c$ states. We utilize $c$ states to uniquely assemble a rectangle of the desired size.

Case 2: $n > 2^{16}$. Let $b = \lceil (\frac{\log n}{\log \log n})^\frac{1}{2} \rceil$, and $d = 4 \lceil \frac{ \log n}{\log \log n}\rceil $. We initialize a variable base counter with value $b^d-n$ represented in base $b$. The binary counter states then attach to this string, counting up to $b^d$, for a total length of $n$. We prove that for $n > 2^{16}$ that $b^d \geq n$. Let $b' =  (\frac{\log n}{\log \log n})^\frac{1}{2} $ and $d' =  \frac{ 4\log n}{\log \log n} $.

\begin{align*} 
 \log(b^{d}) &\geq \log(b'^{d'}) \\
 \log(b^{d}) &\geq d' \log(b') \\
\log(b^{d}) &\geq d' \log \left( \left( \frac{d'}{4} \right) ^\frac{1}{2} \right) \\ 
\log(b^{d}) &\geq \frac{d'}{2} \log \left( \frac{d'}{4} \right) \\ 
\log(b^{d}) &\geq  \frac{ 2\log n}{\log \log n}  \log \left(  \frac{ \log n}{\log \log n} \right)\\ 
\log(b^{d}) &\geq \frac{ 2\log n}{\log \log n} (\log \log n - \log \log \log n)\\ 
\log(b^{d}) &\geq 2\log n - 2\log n \left( \frac{\log \log \log n}{\log \log n} \right)  \\
\end{align*}

Since $n > 2^{16}$, $\frac{\log \log \log n}{\log \log n} < \frac{1}{2}$.

\begin{align*} 
\log(b^{d}) &\geq 2\log n - 2\log n \left( \frac{1}{2} \right) \\
\log(b^{d}) &\geq \log n\\ 
2^{\log(b^{d})} &\geq 2^{\log n}\\ 
b^d &\geq n 
\end{align*}

By Corollary \ref{col:arbBase} we create the assembly that represents the necessary $d$-digit base $b$ string  with $d^\frac{1}{2} + b = \Theta\left( (\frac{\log n}{\log \log n})^\frac{1}{2}\right)$ states. The counter that builds off this string requires $\mathcal{O}(b)$ unique states. Therefore, for all $n > 0$ and a constant $c$, there exists a system that uniquely assembles a $\mathcal{O}(\frac{\log n}{\log \log n}) \times n$ rectangle with $\Theta(b) + \Theta(d^{\frac{1}{2}}) + c = \Theta\left( (\frac{\log n}{\log \log n})^\frac{1}{2}\right)$ states.
\end{proof}

\subsection{Constant Height Rectangles}

In this section, we investigate bounded-height rectangles. We introduce a constant-height counter by first turning the string construction 90 degrees and adding some modifications. For clarity, we change the north and south caps into west and east caps, respectively. We then add states to count down from the number on the string, growing the assembly every time the number is decremented. These states include an additive state $a$, a borrow state $br$ and the attachment state $+$.

\begin{figure}[t]
	\centering
	\begin{subfigure}[b]{.24\textwidth}
		\centering
		\includegraphics[width=.8\textwidth]{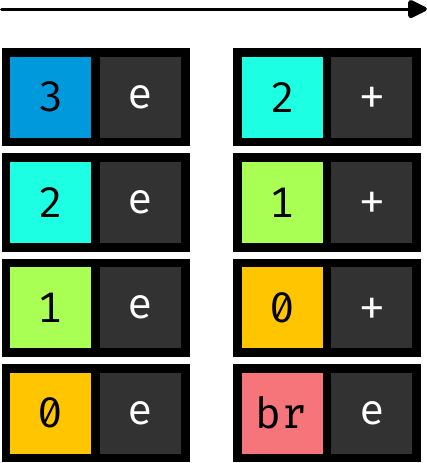}
		\caption{}
		\label{fig:decrementRules}
	\end{subfigure}
	\begin{subfigure}[b]{.24\textwidth}
		\centering
		\includegraphics[width=.9\textwidth]{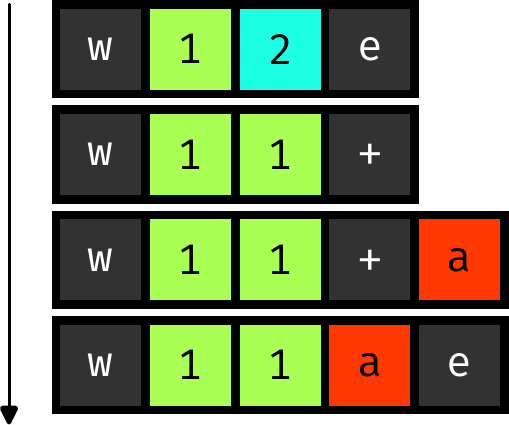}
		\caption{}
		\label{fig:decrementAssembly}
	\end{subfigure}
	\begin{subfigure}[b]{.24\textwidth}
		\centering
		\includegraphics[width=.8\textwidth]{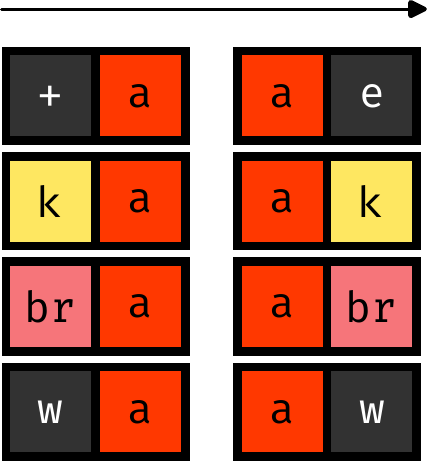}
		\caption{}
		\label{fig:additiveRules}
	\end{subfigure}
	\begin{subfigure}[b]{.24\textwidth}
		\centering
		\includegraphics[width=.9\textwidth]{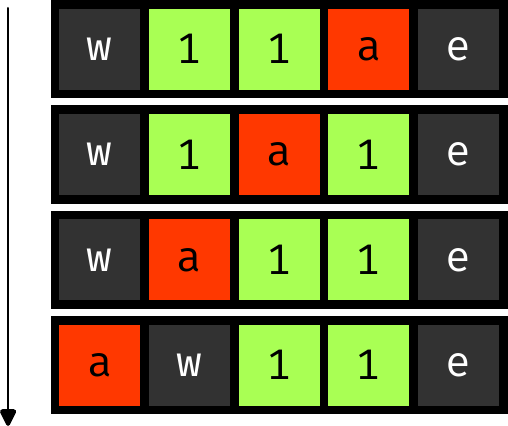}
		\caption{}
		\label{fig:additiveAssembly}
	\end{subfigure}
    \caption{Base 3 number. (a) Transition rules for decrementing. (b) One iteration of the counter. (c) Transition rules for the additive states. The $k$ state can be any digit. (d) The additive state moving to the left of the assembly via transitions. }
    \label{fig:decrement}\label{fig:additive}
\end{figure}


\begin{figure}[t]
	\centering
	\begin{subfigure}[b]{.25\textwidth}
		\centering
		\includegraphics[width=.7\textwidth]{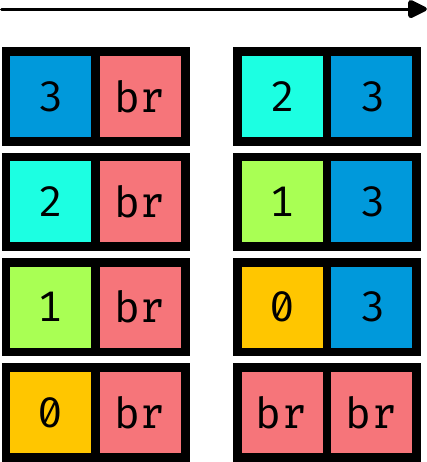}
		\caption{}
		\label{fig:borrowRules}
	\end{subfigure}
	\begin{subfigure}[b]{.25\textwidth}
		\centering
		\includegraphics[width=.9\textwidth]{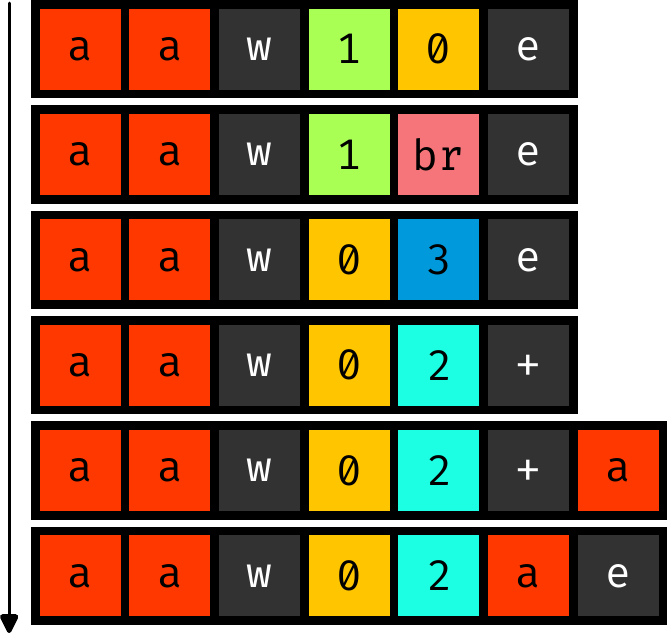}
		\caption{}
		\label{fig:borrowAssembly}
	\end{subfigure}
	\begin{subfigure}[b]{.4\textwidth}
		\centering
	\includegraphics[width=.94\textwidth]{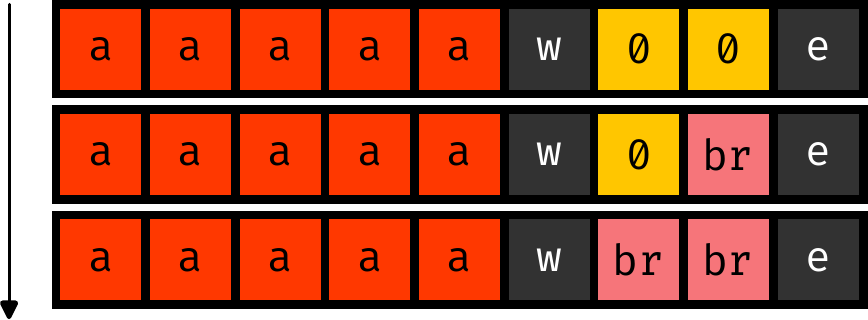}
	\caption{}
	\label{fig:constHeightFinal}
	\end{subfigure}
    \caption{Base 3 number. (a) Transition rules for borrowing. (b) One iteration of the counter with borrowing. (c) The borrow states reaching $w$, stopping the assembly.}
    \label{fig:borrow}
\end{figure}


The constant height counter takes inspiration from long subtraction. The counter begins with trying to subtract 1 from the first place value. An example of this can be seen in Figure \ref{fig:decrement}. If subtraction is possible, the number is decremented and the east tile $e$ will transition into the attachment state $+$ to allow an additive tile to attach. Once attached, the two states $(+, a)$ will transition into $(a, e)$, completing one iteration of the counter. The additive state will eventually make its way to the left side of the assembly as the next iteration begins. An example of this is shown in Figure \ref{fig:additive}. Lastly, if subtraction from the first place value is not possible, meaning the value of the digit is 0, then the system will attempt to borrow from the higher place values. The rules for this as well as an example can be seen in Figure \ref{fig:borrow}. If the borrow state $br$ reaches the west tile $w$, then the number is 0 and the counter is finished. The assembly is complete once the counter has finished and the additive states have all made it to the left side of the west tile $w$. This is shown in Figure \ref{fig:constHeightFinal}. The counter can work with any base $b$ by modifying the transition rules and adding $\mathcal{O}(b)$ states. This counter is not freezing.

\begin{theorem} \label{thm:constHeight}
    For all $n > 0$, there exists a Tile Automata system that uniquely assembles an $4 \times n$ rectangle with nondeterministic transition rules and $\BO{\log ^\frac{1}{4} n}$ states.
\end{theorem}

\begin{proof}
To uniquely assemble a $4 \times n$ rectangle, we construct a constant height counter as described above. By Theorem \ref{thm:nonDetstrings} we construct a binary string with $\BO{\log ^\frac{1}{4} n}$ states. The string is $k$, written in binary, such that $k = n - s$ where $s$ is the number of tiles needed to assemble the string. Once the string has been assembled, the constant height counter can begin counting down from $k$ and attaching tiles. Once the string reaches 0, $k$ tiles have been added to the assembly. The assembly is now $s + k$ tiles long for a total of $n$.
\end{proof}

\subsubsection{Single-Transition Rules}

The constant height counter is easily modified to only use single-transition rules. For every nonsingle-transition rule $\delta$, add one additional state and replace $\delta$ with 3 single-transition rules. For example, given a rule $\delta = (\mathsf{A}, \mathsf{B}, \mathsf{C}, \mathsf{D}, d)$. We add an additional state $\omega$ and three rules as follows.

\begin{itemize}
    \item $\delta_1 = (\mathsf{A}, \mathsf{B},  \mathsf{A}, \omega, d)$
    \item $\delta_2 = (\mathsf{A}, \omega, \mathsf{C}, \omega, d)$
    \item $\delta_3 = (\mathsf{C}, \omega, \mathsf{C}, \mathsf{D}, d)$
\end{itemize}

\begin{theorem} \label{thm:STconstHeight}
    For all $n > 0$, there exists a Tile Automata system that uniquely assembles an $3 \times n$ rectangle with single-transition rules and $\BO{\log ^\frac{1}{3} n}$ states.
\end{theorem}

\begin{proof}
We take the constant height counter and make all rules single-transition using the process described above. Then, by Theorem \ref{thm:SRstrings}, we construct a binary string with $\BO{\log ^\frac{1}{3} n}$ states.
\end{proof}

\subsubsection{Deterministic rules}

\begin{theorem} \label{thm:detConstHeightScaled}
    For all $n > 0$, there exists a deterministic Tile Automata system that uniquely assembles a $2 \times n$ rectangle with single-transition rules and $\Theta\left( (\frac{\log n}{\log \log n})^\frac{1}{2}\right)$ states.
\end{theorem}

\begin{proof}
There are two cases and we use the same ideas as in the proof of Theorem \ref{thm:varBaseDetRec}. Case 1 is the same as in Theorem \ref{thm:varBaseDetRec}.

Case 2: $n > 2^{16}$. Let $b = \lceil (\frac{\log n}{\log \log n})^\frac{1}{2} \rceil$, and $d = 4 \lceil \frac{ \log n}{\log \log n}\rceil $. By Corollary \ref{col:arbBase}, a base-$b$ $d$-digit string can be assembled with $\mathcal{O}(d^{\frac{1}{2}} + b)$ states and single-transition rules at a size of $(n + 2) \times 2$. The initial string will be the value $b^d - n$ represented in base $b$ using $\Theta\left( (\frac{\log n}{\log \log n})^\frac{1}{2}\right)$ states. Since we have an arbitray base, we need to add $\Theta\left( b \right)$ states for the counter to support base $b$. Therefore, for all $n > 0$, there exists a deterministic Tile Automata system that uniquely assembles a $2 \times n$ rectangle with $\Theta\left( b \right) + \Theta\left( d^\frac{1}{2} \right) + c = \Theta\left( (\frac{\log n}{\log \log n})^\frac{1}{2} \right)$ states, where $c$ is a constant from Case 1.
\end{proof}

\subsubsection{Deterministic $1 \times n$ lines}

By using Theorem \ref{thm:lineString} to assemble the initial string of the counter, we achieve $1 \times n$ lines with double transition rules.

\begin{theorem} \label{thm:detConstHeight}
    For all $n > 0$, there exists a deterministic Tile Automata system that uniquely assembles a $1 \times n$ rectangle with $\Theta\left( (\frac{\log n}{\log \log n})^\frac{1}{2}\right)$ states.
\end{theorem}

\begin{proof}
As with the previous proof, let $b = \lceil (\frac{\log n}{\log \log n})^\frac{1}{2} \rceil$, and $d = 4 \lceil \frac{ \log n}{\log \log n}\rceil $. By Theorem \ref{thm:lineString}, a base-$b$ $d$-digit string can be assembled with $\mathcal{O}(d^{\frac{1}{2}} + b)$ states at exact scale. $\Theta \left( b \right)$ states are added for the counter to support the arbitary base $b$. Therefore, for all $n > 0$, there exists a deterministic Tile Automata system that uniquely assembles a $1 \times n$ rectangle with $\Theta\left( b \right) + \Theta\left( d^\frac{1}{2} \right) + c = \Theta\left( (\frac{\log n}{\log \log n})^\frac{1}{2} \right)$ states, where $c$ is a constant from Case 1.
\end{proof}

\section{Squares}\label{sec:squares}
In this section we utilize the rectangle constructions to build $n \times n$ squares using the optimal number of states.

Let $n' = n - 4\lceil \frac{\log n}{\log \log n} \rceil - 2$, and $\Gamma_0$ be a deterministic Tile Automata system that builds a $n' \times (4\lceil \frac{\log n}{\log \log n} \rceil + 2)$  rectangle using the process described in Theorem \ref{thm:varBaseDetRec}. Let $\Gamma_1$ be a copy of $\Gamma_0$ with the affinity and transition rules rotated $90$ degrees clockwise, and the state labels appended with the symbol ``*1''. This system will have distinct states from $\Gamma_0$, and will build an equivalent rectangle rotated $90$ degrees clockwise. We create two more copies of $\Gamma_0$ ($\Gamma_2$ and $\Gamma_3$), and rotate them $180$ and $270$ degrees, respectively. We append the state labels of $\Gamma_2$ and $\Gamma_3$ in a similar way.

We utilize the four systems described above to build a hollow border consisting of the four rectangles, and then adding additional initial states which fill in this border, creating the $n \times n$ square.

We create $\Gamma_n$, starting with system $\Gamma_0$, and adding all the states, initial states, affinity rules, and transition rules from the other systems ($\Gamma_1, \Gamma_2, \Gamma_3$). The seed states of the other systems are added as initial states to $\Gamma_n$. We add a constant number of additional states and transition rules so that the completion of one rectangle allows for the ``seeding'' of the next.

\begin{figure}[t]
	\centering
	\includegraphics[width=1.\textwidth]{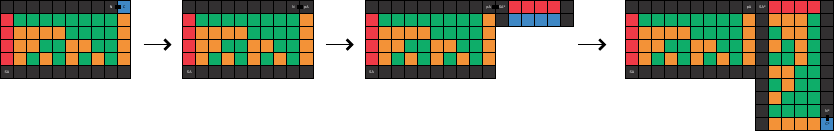}
	\caption{The transitions that take place after the first rectangle is built. The carry state transitions to a new state that allows a seed row for the second rectangle to begin growth}
	\label{fig:square1}
\end{figure}

\textbf{Reseeding the Next Rectangle.}
To $\Gamma_n$ we add transition rules such that once the first rectangle (originally built by $\Gamma_0$) has built to its final width, a tile on the rightmost column of the rectangle will transition to a new state $pA$. $pA$ has affinity with the state $S_A*1$, which originally was the seed state of $\Gamma_1$. This allows state $S_A*1$ to attach to the right side of the rectangle, ``seeding'' $\Gamma_1$  and allowing the next rectangle to assemble (Figure \ref{fig:square1}). The same technique is used to seed $\Gamma_2$ and $\Gamma_3$.

\begin{figure}[t]
	\centering
	\includegraphics[width=1.\textwidth]{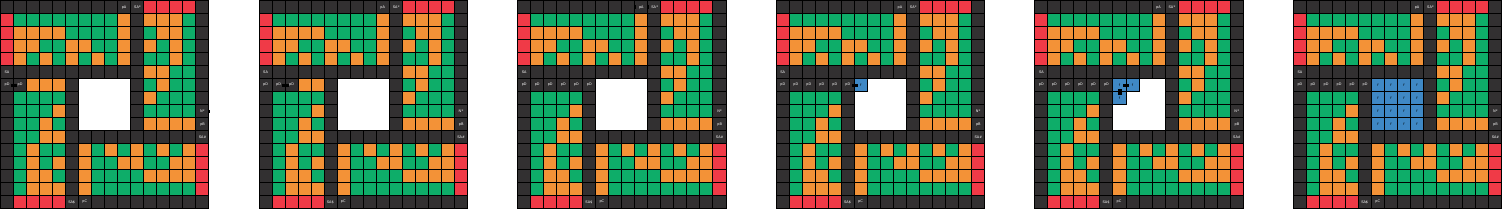}
	\caption{Once all 4 sides of the square build the $pD$ state propagates to the center and allows the light blue tiles to fill in}
	\label{fig:square2}
\end{figure}

\textbf{Filler Tiles.}
When the construction of the final rectangle (of $\Gamma_3$) completes, transition rules propagate a state $pD$ towards the center of the square (Figure \ref{fig:square2}). Additionally, we add an initial state $r$, which has affinity with itself in every orientation, as will as with state $pD$ on its west side. This allows the center of the square to be filled with tiles.

\begin{theorem}\label{thm:SQ}
For all $n > 0$, there exists a Tile Automata system that uniquely assembles an $n \times n$ square with,

\begin{itemize}
	\item Deterministic transition rules and $\Theta\left((\frac{\log n}{\log \log n})^\frac{1}{2}\right)$ states.
	\item Single-Transition rules and  $\Theta(\log ^\frac{1}{3} n)$ states.
	\item Nondeterministic transition rules and $\Theta(\log ^\frac{1}{4} n)$ states.
\end{itemize}
\end{theorem}

%
%
%
%
%

\section{Future Work} \label{sec:future}
This paper showed optimal bounds for uniquely building $n \times n$ squares in three variants of seeded Tile Automata without cooperative binding. En route, we proved optimal bounds for constructing strings and rectangles. Serving as a preliminary investigation into constructing shapes in this model. This leaves many open questions:    

\begin{itemize}
	\item We achieve building a $1 \times n$ line using with deterministic rules with $\Theta( (\frac{\log n}{\log \log n})^\frac{1}{2} )$ states. Is it possible to achieve the optimal bounds for nondeterministic rules? 
	\item We allow transition rules between non-bonded tiles. 
	 Can the same results be achieved with the restriction that a transition rule can only exist between two tiles if they share an affinity in the same direction?
	\item While we show optimal bounds can be achieved without cooperative binding, can we simulate so-called zig-zag aTAM systems? These are a restricted version of the cooperative aTAM that is capable of Turing computation.
	\item We show efficient bounds for constructing strings in Tile Automata. Given the power of the model, it should be possible to build algorithmically defined shapes such as in \cite{soloveichik2007complexity} by printing Komolgorov optimal strings and inputting them to a Turing machine.  
	\item Surface chemical reaction networks (sCRN) is another model of asynchronous cellular automata. A key difference between this model and Tile Automata is sCRN transition rules (reactions) do not have a direction and are written of the form $A + B \rightarrow C + D$.   This means anytime the species/state $A$ is adjacent to $B$, then they change to $C$ and $D$, respectively.   
\end{itemize}

\bibliographystyle{plain}
\bibliography{taSquareBib}

\end{document}